\documentclass[aos,preprint]{imsart}

\setattribute{journal}{name}{}

\RequirePackage[OT1]{fontenc}
\RequirePackage{amsthm,amsmath, amsfonts,amssymb,mathrsfs, bbm, bm}
\RequirePackage[colorlinks,citecolor=blue,urlcolor=blue]{hyperref}
\usepackage{algorithm, algpseudocode}
\usepackage{graphicx, epstopdf, subcaption}
\usepackage{marginnote}
\usepackage{multirow}
\usepackage{algorithm}

\usepackage[numbers]{natbib}
\startlocaldefs
\numberwithin{equation}{section}
\theoremstyle{plain}
\newtheorem{theorem}{Theorem}[section]
\newtheorem{corollary}[theorem]{Corollary}

\newtheorem{proposition}[theorem]{Proposition}

\endlocaldefs

\theoremstyle{definition}
\newtheorem{example}{Example}[section]
\newtheorem{remark}{Remark}[section]
\newtheorem{assumption}{Assumption}[section]

\usepackage{tikz}

\newcommand{\E}{\mathbb{E}}

\def\bp{{\bf{p}}}

\def\bnu{{\bf{\nu}}}
\def\bw{{\bf{w}}}

\def\bu{{\bf{u}}}

\def\bsvarep{{\boldsymbol{\varepsilon}}}

\def\HH{\mathcal H}

\def\WW{\mathcal W}

\def\BB{\mathcal B}

\def\NN{\mathcal N}

\def\E{\mathbb E}

\def\N{\mathbb N}

\def\R{\mathbb R}

\def\P{\mathbb P}

\usepackage{xr}
\begin{document}

\begin{frontmatter}
\title{Another look at Statistical Calibration: a Non-Asymptotic Theory and Prediction-Oriented  Optimality}
\runtitle{Another look at Statistical Calibration}
\begin{aug}
\author{\fnms{Xiaowu} \snm{Dai}\thanksref{t1, t2}
\ead[label=e1]{xdai26@wisc.edu}}
\and
\author{\fnms{Peter} \snm{Chien}\thanksref{t2}
\ead[label=e2]{	peter.chien@wisc.edu}}

\thankstext{t1}{Supported in part by NSF Grant DMS-1308877.}
\thankstext{t2}{Supported in part by NSF Grant DMS-1564376.}
\runauthor{X. Dai and P. Chien}

\affiliation{University of Wisconsin-Madison}

\address{Department of Statistics\\
University of Wisconsin-Madison\\
1300 University Avenue\\
Madison, Wisconsin 53706\\
USA\\
\printead{e1}\\
\phantom{E-mail:\ }\printead*{e2}}
\end{aug}

\begin{abstract}
We provide another look at the statistical calibration problem in computer models.
This viewpoint is inspired by two overarching practical considerations of computer models: (i) many computer models are inadequate for perfectly modeling physical systems, even with the best-tuned calibration parameters; (ii) only a finite number of data points are available from the physical experiment associated with a computer model.
Following this new line of thinking, we provide a non-asymptotic theory and derive a prediction-oriented calibration method.
Our calibration method  minimizes the predictive mean squared error for a finite sample size with statistical guarantees.
We introduce an algorithm to perform the proposed calibration method and connect it to existing Bayesian calibration methods.
Synthetic and real examples are provided to corroborate the derived theory and illustrate some advantages of the proposed calibration method.
\end{abstract}



\end{frontmatter}

\section{Introduction}
\label{sec:intro}
In engineering and sciences, computer models are increasingly used for studying complex physical systems such as those in cosmology, weather forecasting, material science, and shock physics  \cite{santner2013design}.
Let $Y$ denote the output from a physical system $\zeta(\cdot)$ with a set of inputs $X$.
For $i= 1,\ldots, n$, assume $Y_i$ has the following form:
\begin{equation}
\label{eqn:phyobs}
Y_i = \zeta(X_i) + \varepsilon_i,
\end{equation}
where the random error $\varepsilon_i$ follows the independent $N(0,\sigma^2)$ distribution and the design points $X_i$ has support on $\Omega=[0,1]^d$.
Let $\eta(x,\theta)$ denote a computer model for approximating $\zeta(x)$ with inputs $x\in\Omega$ and \emph{calibration parameters} $\theta\in\Theta\subset \R^p$.
The values of $\theta$ cannot be directly measured and typically unknown in the physical data.
As George Box famously stated \emph{``All models are wrong, but essentially some are useful"}, even the best computer models are only approximations of reality.
It is possible to enhance the quality of the computer model
$\eta(x,\theta)$ by tuning or calibrating the calibration parameters $\theta$. 
But in most practical scenarios, the computer output $\eta(x,\theta)$ cannot fit the physical response $\zeta(x)$ perfectly,
regardless how the calibration parameters $\theta$ are best tuned
\cite{kennedy2001bayesian, santner2013design}.

Another practical fact is only a \emph{finite} number, $n$, of data points are available from the physical experiment in (\ref{eqn:phyobs}) to tune $\theta$.

By simultaneously following these two practical considerations, we take a new look at the calibration problem.
Our purpose is to optimally predict $\zeta(\cdot)$ by calibrating $\theta$ in $\eta(\cdot,\theta)$ and estimating the model discrepancy $\zeta(x) - \eta(x,\theta)$ based on a finite number of physical data. To this end, we use nonparametric approaches to modeling the physical system and the discrepancy function. We establish a non-asymptotic minimax  estimation risk for nonparametric regression and achieve the optimal risk by using regularized estimators in the  reproducing kernel Hilbert space (RKHS) \cite{aronszajn1950theory, wahba1990}.
We show that our problem of prediction oriented calibration is equivalent to finding the minimizer of the model discrepancy under the RKHS norm.  We further establish an exact statistical guarantee in the sense that for a finite sample of physical observations, the prediction error is minimized by using the computer model calibrated with the proposed  method. Furthermore, we provide an algorithm to estimate the optimal calibration parameters and the model  discrepancy.

\subsection{Comparison with existing work}

We discuss the differences between our frequentist calibration method and other frequentist calibration methods in the literature. 
\cite{joseph2009statistical} considers calibration using a parametric form of discrepancy. 
We use a nonparametric approach to better modeling the physical system and the model discrepancy.
The $L_2$-calibration  method in \cite{tuo2015efficient} minimizes
the model discrepancy under  the $L_2$-norm.  \cite{wong2017frequentist} performs calibration by minimizing the model discrepancy under the empirical $l_2$-norm.
Below are the main differences between  \cite{tuo2015efficient, wong2017frequentist} and ours.
\begin{itemize}
\item Different method. Calibration in \cite{tuo2015efficient, wong2017frequentist} and our work minimizes the following different norms of the model discrepancy: the $L_2$-norm in \cite{tuo2015efficient}, the empirical $l_2$-norm in  \cite{wong2017frequentist} and the RKHS norm in our method.
\item Different analysis. Theoretical results in \cite{tuo2015efficient, wong2017frequentist}  are based on asymptotics theory assuming the number of physical observations go to infinity. Our theory is based on finite-sample properties of calibration and prediction following the fact that usually, only a finite number of  physical data are available.
\item Different results. The $L_2$-calibration in \cite{tuo2015efficient} minimizes the distance between the physical system and the imperfect computer model, but not directly for predicting the physical system. 
    \cite{wong2017frequentist} performs the least square calibration 
    and then estimates the model discrepancy  in the RKHS. However, for a finite number of physical observations, 
      the estimation error of  discrepancy can be large. To overcome this difficulty, our calibration method minimizes the  predictive mean squared error  for a finite sample of physical data with statistical guarantees.
\end{itemize}

Bayesian calibration was studied by \cite{kennedy2001bayesian, oakley2004probabilistic, higdon2004combining, higdon2008computer,  joseph2015engineering, plumlee2017bayesian, tuo2018prediction}, among others. Our frequentist calibration method is easier to compute and complements these Bayesian  methods. Furthermore, we will discuss an interesting connection between our  method and these Bayesian methods in Section \ref{sec:algorithm}. This link offers new justification for 
observed successful  prediction performance of the Bayesian calibration methods in various fields. 

Our non-asymptotic minimax theory is inspired by recent developments of  concentration inequalities that provide valid statistical inference and estimation results for  finite samples. Existing research on concentration inequalities 
typically addresses finite dimensional parameters for parametric models. Because our interest is computer model calibration, we develop a non-asymptotic minimax theory for nonparametric models.

The remainder of the article is organized as follows. In Section \ref{sec:optimalcalibration}, we discuss the identifiability issue and formulate a prediction-oriented optimal calibration method. In Section \ref{sec:finitesampleproperies}, we establish a non-asymptotic minimax theory and apply it to the prediction-oriented calibration method. In Section \ref{sec:algorithm}, we develop an algorithm for solving our calibration problem and build a connection between our method and the Bayesian calibration method.
In Section \ref{sec:simulationandareal}, we provide synthetic and real examples to corroborate the derived theory and illustrate some advantages of the proposed calibration method. We give concluding remarks in Section \ref{sec:concludingremarks}. All other proofs and algorithm are delegated to the supplementary material.

\section{Prediction-oriented calibration}
\label{sec:optimalcalibration}
Since the computer model is imperfect for modeling the physical system,  $\eta(x,\theta) \not\equiv\zeta(x)$ for any $\theta\in\Theta$.
A model discrepancy function $\delta(x,\theta) = \zeta(x) - \eta(x,\theta)$ is commonly used \cite{kennedy2001bayesian, martins2008class}.  Equivalently, write
\begin{equation}
\label{eqn:discrepancymodel}
\zeta(x) = \eta(x,\theta) +  \delta(x,\theta), \ \  \forall x\in\Omega, \theta\in\Theta.
\end{equation}
The goal is to accurately predict $\zeta(\cdot)$ using the computer model, which requires calibrating $\theta$ and estimating $\delta(\cdot,\theta)$  simultaneously with a finite physical sample in (\ref{eqn:phyobs}).
Two main difficulties arise. The identifiability issue of $\theta$ to be discussed in Section \ref{sec:identifability}  and the non-negligible estimation error of $\delta(\cdot,\theta)$ due to the finite sample of the physical data. These two issues motivate our calibration method in Section \ref{sec:defofoptcal}.

\subsection{The identifiability issue}
\label{sec:identifability}

Suppose that  $\zeta(\cdot)$ resides in a RKHS $(\HH,\|\cdot\|_\HH)$ on $\Omega=[0,1]^d$. One example of $\HH$ is the $m$th  order Sobolev space $\WW_2^m(\Omega)$ with $2m>d$:
\begin{equation*}
\begin{aligned}
\WW_2^m(\Omega) & = \left\{\left.g(\cdot)\in L_2(\Omega)\vphantom{\frac{a}{b}}\right|\frac{\partial^{\alpha_1+\cdots+\alpha_d} }{\partial^{\alpha_1}x_1\cdots\partial^{\alpha_d}x_d}g(\cdot) \in L_2(\Omega),\right.\\
& \quad\quad \quad\quad\quad\quad\quad\left. \forall \alpha_1,\ldots\alpha_d\in\N \text{ with }\alpha_1+\cdots+\alpha_d\leq m \vphantom{\frac{\partial^{\alpha_1+\cdots+\alpha_d} }{\partial^{\alpha_1}x_1\cdots\partial^{\alpha_d}x_d}}\right\}.
\end{aligned}
\end{equation*}
Since $\eta(\cdot,\theta)$ approximates the physical system $\zeta(\cdot)$, we assume the following regularity condition for $\eta(\cdot,\theta)$.
\begin{assumption}
\label{asp:etaspace}
For any $\theta\in\Theta$, the computer model $\eta(\cdot,\theta)\in\HH$.
\end{assumption}
An analogy of Assumption \ref{asp:etaspace} was already used in \cite{plumlee2017bayesian}. But instead of assuming the function space is a RKHS, \cite{plumlee2017bayesian} considers  a function space of bounded  mixed derivatives.
Assumption \ref{asp:etaspace} implies that for any $\theta\in\Theta$, $\delta(\cdot,\theta) = \zeta(x) - \eta(x,\theta)\in\HH$. This observation leads to a potential identifiability issue for $\theta$.
For example, for two different $\theta_1\neq \theta_2\in\Theta$,  their corresponding model discrepancies  $\delta(x,\theta_1) = \zeta(x) - \eta(x,\theta_1)$ and $\delta(x,\theta_2) = \zeta(x) - \eta(x,\theta_2)$ are both in $\HH$.
This implies that both $(\theta_1,\delta(\cdot,\theta_1)), (\theta_2,\delta(\cdot,\theta_2))\in \Theta\times\HH$ are true  for model (\ref{eqn:discrepancymodel}). There are  infinite truth pairs $(\theta,\delta(\cdot,\theta))\in\Theta\times\HH$ for (\ref{eqn:discrepancymodel}) by choosing arbitrary $\theta\in\Theta$ and using $\delta(\cdot,\theta)=\zeta(x) - \eta(x,\theta)$.
This identifiability issue was first noticed by K. Beven and P. Diggle in their discussion of \cite{kennedy2001bayesian}.

If Assumption \ref{asp:etaspace} fails, our results hold for replacing $\delta(\cdot,\theta) = \zeta(\cdot)-\eta(\cdot,\theta)$ with $\delta(\cdot,\theta) = P_\HH\{\zeta(\cdot)-\eta(\cdot,\theta)\}$, where $P_\HH$ denotes the projection from $L_2(\Omega)$ to $\HH$ in  the $L_2$-norm.
Hereinafter, we focus on the imperfect computer model under Assumption \ref{asp:etaspace}.

\subsection{Definition of prediction-oriented calibration}
\label{sec:defofoptcal}
Denote by $\Pi$ the sampling distribution of $X_i$ in (\ref{eqn:phyobs}) which is independent of $\varepsilon_i$ and  satisfies $\Pi(\Omega)=1$. Assume that the density of $\Pi$ is bounded away from zero and infinity on $\Omega$.
Let $X^*$ be drawn from $\Pi$ and $Y^* = \zeta(X^*) + \varepsilon^* = \eta(X^*,\theta) + \delta(X^*,\theta)+ \varepsilon^*$ with $\varepsilon^*\sim N(0,\sigma^2)$. Then for a fixed $\theta\in\Theta$, the minimal predictive mean squared error for predicting $Y^*$ is
\begin{equation}
\label{eqn:minimalmspred}
\begin{aligned}
&  \inf_{\widetilde{\delta}}\E\left\{Y^* - [\eta(X^*,\theta) + \widetilde{\delta}(X^*,\theta)]\right\}^2\\
&\ \ \ \  =  \sigma^2 + \inf_{\widetilde{\delta}}\|\delta(\cdot,\theta) - \widetilde{\delta}(\cdot,\theta)\|_{L_2(\Pi)}^2,
\end{aligned}
\end{equation}
where the infimum is over all measurable estimators.

The identifiability issue discussed in Section \ref{sec:identifability} indicates that there are infinite pairs of $(\theta,\delta(\cdot,\theta))\in\Theta\times\HH$ satisfying model (\ref{eqn:discrepancymodel}). However, for a finite sample size $n$, the minimal estimation error $\inf_{\widetilde{\delta}}\|\delta(\cdot,\theta) - \widetilde{\delta}(\cdot,\theta)\|_{L_2(\Pi)}$ does not vanish \cite{cover2012elements}.
This fact motivates us to define \textit{optimal calibration} $\theta^{\text{opt-pred}}$ to minimize the minimal predictive mean squared error (\ref{eqn:minimalmspred}) over $\theta\in\Theta$.  Equivalently, we define
\begin{equation}
\label{def:btheta0}
\theta^{\text{opt-pred}}  \equiv  \underset{\theta\in\Theta}{\arg\min} \left\{\inf_{\widetilde{\delta}}\|\delta(\cdot,\theta) - \widetilde{\delta}(\cdot,\theta)\|_{L_2(\Pi)}\right\},
\end{equation}
where the superscript ``opt-pred" denotes ``optimal for prediction".
By definition, $\theta^{\text{opt-pred}}$  in (\ref{def:btheta0}) is optimal for predicting $\zeta(\cdot)$.
We will  establish a non-asymptotic theory for nonparametric regression in Section \ref{sec:finitesampleproperies} and introduce an algorithm in Section \ref{sec:algorithm}.
The formulation of $\theta^{\text{opt-pred}}$  in (\ref{def:btheta0}) is frequentist. 
We will discuss in Section \ref{sec:comparewithfrequentist} the differences between $\theta^{\text{opt-pred}}$ and other frequentist calibration methods, including the $L_2$-calibration method \cite{tuo2015efficient} and the least square calibration method \cite{wong2017frequentist}.

\section{Main results}
\label{sec:finitesampleproperies}

Existing theory for calibration, including \cite{tuo2015efficient, wong2017frequentist}, adopts the asymptotic regime that the number of data points for the physical experiment goes to infinity. 
In Section \ref{sec:nonasymresfornonpara}, we present non-asymptotic minimax  risk for nonparametric models and apply it the model calibration problem in Section \ref{sec:apptooptcalibration}. 
In Section \ref{sec:improvbycompmodels}, we show the improvement in prediction achieved by incorporating data from computer models. In Section \ref{sec:comparewithfrequentist}, we compare our calibration method with some existing frequentist methods.

\subsection{Non-asymptotic minimax theory for nonparametric regressions}
\label{sec:nonasymresfornonpara}

We consider the nonparametric regression in (\ref{eqn:phyobs}), where $\zeta(\cdot)$ resides in the RKHS
$\HH$. Let $\|\cdot\|_{\HH}$ be the corresponding RKHS norm and  $K:\Omega\times\Omega\to \R$ be a Mercer kernel generating  $(\HH,\|\cdot\|_{\HH})$. By the spectral theorem, $K$ admits the eigenvalue decomposition:
\begin{equation*}
K(x,x') = \sum_{\nu\geq 1}\lambda_\nu\phi_{\nu}(x)\phi_\nu(x'),
\end{equation*}
where $\lambda_1\geq \lambda_2\geq \cdots\geq 0$ are the eigenvalues and $\{\phi_\nu:\nu\geq 1\}$ are the corresponding eigenfunctions such that $\langle\phi_\nu,\phi_{\nu'}\rangle_{L_2(\Pi)} = \delta_{\nu\nu'}$. Here, $\delta_{\nu\nu'}$ is the Kronecker delta.
We assume the polynomial  decay rate of  eigenvalues in Assumption \ref{asp:decayrateofeigen}.
This common assumption holds for the Sobolev space $\HH= \WW_2^m(\Omega)$ with Lebesgue measure on $\Omega$.
\begin{assumption}
\label{asp:decayrateofeigen}
For $2m>d$, suppose that for any $\nu\geq 1$, the eigenvalues satisfy
 $c_\lambda\nu^{-2m/d}\leq\lambda_\nu\leq C_\lambda\nu^{-2m/d}$ with constants $0<c_\lambda<C_\lambda<\infty$, and the eigenfunctions are uniformly bounded: $\max_{x\in\Omega}|\phi_\nu(x)|\leq c_\phi$ with a constant  $c_\phi$ for any $\nu\geq 1$.
\end{assumption}

We now present our main results. We first show a non-asymptotic minimax lower bound for the predictive  mean squared error.
\begin{theorem}
\label{thm:lowerbdnonasymnonpara}
Under the regression model (\ref{eqn:phyobs}) where $\zeta(\cdot)\in\HH$ and Assumption \ref{asp:decayrateofeigen} holds, there exists a constant $C_1>0$  not depending on  $n, \sigma, m, d, \|\zeta\|_\HH$   such that for any $n\geq 1$,
 \begin{equation*}
\begin{aligned}
&  \inf_{\widetilde{\zeta}}\sup_{\zeta\in\HH}\P\left\{\|\widetilde{\zeta}- \zeta\|_{L_2(\Pi)}^2 \geq C_1 \left[1+n^{-\frac{2m-d}{4m+2d}}\left(1+{\sigma}/{\|\zeta\|_\HH}\right)^{-\frac{2d}{2m+d}}\right]^2\right.\\
& \quad\quad\quad\quad\quad\quad\quad\quad\quad\quad\left.\cdot n^{-\frac{2m}{2m+d}}\left(\|\zeta\|_\HH+\sigma\right)^2\left(1+{\sigma}/{\|\zeta\|_\HH}\right)^{-\frac{2d}{2m+d}}\right\}>0.
\end{aligned}
\end{equation*}
\end{theorem}
The proof of this theorem, given in the supplementary material, is based on Fano's lemma \cite{cover2012elements}. Now, we show that the non-asymptotic  lower bound of the theorem can be achieved by the regularized estimator in the RKHS:
\begin{equation}
\label{eqn:regestforzeta}
\widehat{\zeta}_{n\lambda} = \underset{g\in \HH}{\arg\min}\left\{\frac{1}{n}\sum_{i=1}^n[Y_i - g(X_i)]^2 + \lambda\|g\|^2_{\HH}\right\},
\end{equation}
where $\lambda> 0$ is the tuning parameter. The following result shows that $\widehat{\zeta}_{n\lambda} $ is minimax optimal for a finite sample.

\begin{theorem}
\label{thm:hatfferror}
Under the regression model (\ref{eqn:phyobs}) where $\zeta(\cdot)\in\HH$ and Assumption \ref{asp:decayrateofeigen}, there exists a constant $C_2>0$  not depending on $n, \sigma, m, d, \|\zeta\|_\HH$  such that for any $n\geq 1$
and any $\alpha> 0$ with probability at least $1-8e^{-\alpha^2}$,
\begin{equation*}
\begin{aligned}
&  \|\widehat{\zeta}_{n\lambda}- \zeta\|_{L_2(\Pi)}^2 \leq C_2\left[1+\alpha^{\frac{2m-d}{2m+d}}n^{-\frac{2m-d}{4m+2d}}\left(1+{\sigma}/{\|\zeta\|_\HH}\right)^{-\frac{2d}{2m+d}}\right]^2\\
& \quad\quad\quad\quad\quad\quad\quad\quad\quad\cdot\alpha^{\frac{4m}{2m+d}}n^{-\frac{2m}{2m+d}}\left(\|\zeta\|_\HH+\sigma\right)^2\left(1+{\sigma}/{\|\zeta\|_\HH}\right)^{-\frac{2d}{2m+d}},
\end{aligned}
\end{equation*}
where $\widehat{\zeta}_{n\lambda}$ is defined by (\ref{eqn:regestforzeta}) with $\lambda = n^{-{2m}/{(2m+d)}}\{4m^{1/2}(4m-d)^{-1/2}C_\lambda^{{d}/{4m}}\cdot[\alpha A + \sigma c_\phi(1+\sqrt{2}\alpha)/\|\zeta\|_\HH]\}^{{4m}/{(2m+d)}}$. Here, $A$ is a constant not depending on $n,\sigma,\|\zeta\|_\HH$.
\end{theorem}

We relegate the proof of Theorem \ref{thm:hatfferror} to the supplementary material.
We establish the proof by using results from empirical processes such as the maximal  inequalities and the concentration inequalities  \cite{kosorok2007introduction} and  deriving some new techniques.

We make three remarks on this theorem. First, the conventional
asymptotic convergence rate can be recovered from Theorem \ref{thm:hatfferror}. Since  $2m>d$ in Assumption \ref{asp:decayrateofeigen},
\begin{equation*}
\alpha^{\frac{2m-d}{2m+d}}n^{-\frac{2m-d}{4m+2d}}\left(1+\sigma/\|\zeta\|_\HH\right)^{-\frac{2d}{2m+d}} = o(1),\quad \text{as }n\to\infty.
\end{equation*}
Thus, Theorem \ref{thm:hatfferror} yields that $\|\widehat{\zeta}_{n\lambda}- \zeta\|_{L_2(\Pi)}^2 = O_\P\{n^{-2m/(2m+d)}\}$ as $n\to \infty$. This is a well-known  rate (\cite{cox1984multivariate, wahba1979convergence,  wahba1990}).

Second, Theorems \ref{thm:lowerbdnonasymnonpara} and \ref{thm:hatfferror} together immediate imply that the non-asymptotic minimax optimal risk for estimating $\zeta\in\HH$ is
\begin{equation}
\label{eqn:minoptnonp}
\begin{aligned}
\|\widetilde{\zeta}- \zeta\|_{L_2(\Pi)}^2 & = C \left[1+n^{-\frac{2m-d}{4m+2d}}\left(1+{\sigma}/{\|\zeta\|_\HH}\right)^{-\frac{2d}{2m+d}}\right]^2\\
& \quad\quad\quad\cdot n^{-\frac{2m}{2m+d}}\left(\|\zeta\|_\HH+\sigma\right)^2\left(1+{\sigma}/{\|\zeta\|_\HH}\right)^{-\frac{2d}{2m+d}},
\end{aligned}
\end{equation}
for some constant $C$.
This minimax risk holds for a finite sample and indicates the dependence on the signal-to-noise ratio: $\|\zeta\|_\HH/\sigma$ and the magnitude of signal and noise: $\|\zeta\|_\HH+\sigma$.

Third, Theorem \ref{thm:hatfferror}  suggests a tradeoff between the approximation error and the prediction error.
A smaller $\lambda$ in (\ref{eqn:regestforzeta}) corresponds to a smaller approximation error:
$\frac{1}{n}\sum_{i=1}^n[Y_i - \widehat{\zeta}_{n\lambda}(X_i)]^2$; see, e.g., \cite{wahba1990}.
In practice, we usually do not know the values of $\|\zeta\|_\HH$ or $\sigma$.
Theorem \ref{thm:hatfferror} implies that for achieving the minimax optimal risk (\ref{eqn:minoptnonp}),
the optimal $\lambda$ with a smaller value corresponds to
a larger $\|\zeta\|_\HH$, which yields a larger risk for predction in (\ref{eqn:minoptnonp}).

\subsection{Optimal calibration and prediction}
\label{sec:apptooptcalibration}

We apply the non-asymptotic minimax theory  in Section \ref{sec:nonasymresfornonpara} to
derive an equivalent form for the calibration method in (\ref{def:btheta0}). The following theorem  leads to an algorithm for computing $\theta^{\text{opt-pred}}$ in Section \ref{sec:algorithm}.
\begin{theorem}
\label{thm:thetastardeltahh}
Under the Assumptions in \ref{asp:etaspace} and \ref{asp:decayrateofeigen}, for any $n\geq 1$, the calibration procedure in (\ref{def:btheta0}) is equivalent to finding the minimizer of the model discrepancy equipped with the RKHS norm:
\begin{equation*}
\begin{aligned}
\theta^{\text{opt-pred}} &=  \underset{\theta\in\Theta}{\arg\min} \left\{\|\zeta(\cdot) - \eta(\cdot,\theta)\|_{\HH}\right\}.
\end{aligned}
\end{equation*}
\end{theorem}
The proof of this theorem is relegated to the supplementary material.
Here are some explanations. If the model discrepancy  $\zeta(\cdot) - \eta(\cdot,\theta)$ has a small RKHS norm,  the  discrepancy is a simple function in the RKHS. Since the sample size $n$  is fixed and finite, a simpler function  should have a more accurate estimator by using limited information from data.
As discussed in Section \ref{sec:optimalcalibration}, a more accurate discrepancy estimator gives a smaller prediction error for the physical system.
Unlike the existing work in model calibrations,
Theorem \ref{thm:thetastardeltahh} justifies the use of RKHS norm to measure the model discrepancy, which is  different from the $L_2$-norm  \cite{tuo2015efficient} and the empirical $l_2$-norm  \cite{wong2017frequentist}.

We now discuss the non-asymptotic minimax risk for predicting  $\zeta(\cdot)$ based on the calibration parameters $\theta^{\text{opt-pred}}$. As a corollary of Theorem \ref{thm:lowerbdnonasymnonpara}, the following result gives a lower bound of the predictive mean squared error when using the computer model  calibrated by $\theta^{\text{opt-pred}}$ and an estimator of the model discrepancy.
\begin{corollary}
\label{cor:lowerbd}
 Under the regression model (\ref{eqn:phyobs}) where $\zeta(\cdot)\in\HH$ and  Assumption \ref{asp:etaspace} and \ref{asp:decayrateofeigen} hold, then for any $n\geq 1$,
 \begin{equation*}
\begin{aligned}
&  \inf_{\widetilde{\delta}}\sup_{\zeta\in\HH}\P
\vphantom{\left(1+{\sigma}/{\|\zeta-\eta(\cdot,\theta^{\text{opt-pred}})\|_\HH}\right)^{-\frac{2d}{2m+d}}}
\left\{\|[\eta(\cdot,\theta^{\text{opt-pred}})+\widetilde{\delta}(\cdot)]- \zeta(\cdot)\|_{L_2(\Pi)}^2\right. \\
&  \left.\geq C_1\left[1+n^{-\frac{2m-d}{4m+2d}}\left(1+{\sigma}/{\|\zeta-\eta(\cdot,\theta^{\text{opt-pred}})\|_\HH}\right)^{-\frac{2d}{2m+d}}\right]^2n^{-\frac{2m}{2m+d}} \right.\\
&\quad\left.\cdot \left(\|\zeta-\eta(\cdot,\theta^{\text{opt-pred}})\|_\HH+\sigma\right)^2\left(1+{\sigma}/{\|\zeta-\eta(\cdot,\theta^{\text{opt-pred}})\|_\HH}\right)^{-\frac{2d}{2m+d}}\right\}>0.
\end{aligned}
\end{equation*}
Here, the constant $C_1$ is the same as in Theorem \ref{thm:lowerbdnonasymnonpara} that does not depend on $n, \sigma, m, d, \|\zeta-\eta(\cdot, \theta^{\text{opt-pred}})\|_\HH$.
\end{corollary}

As a corollary of Theorem \ref{thm:hatfferror},  we show that the non-asymptotic  lower bound in Corollary \ref{cor:lowerbd} can be achieved by  the regularized estimator for the model discrepancy in the RKHS $\HH$:
\begin{equation}
\label{eqn:widehatdeltanlambda}
\widehat{\delta}_{n\lambda}(\cdot,\theta) = \underset{h\in\HH}{\arg\min} \left\{\frac{1}{n}\sum_{i=1}^n[Y_i - \eta(X_i,\theta) - h(X_i)]^2 + \lambda\|h\|_{\HH}^2\right\},
\end{equation}
where $\lambda> 0$ is the tuning parameter.
\begin{corollary}
\label{cor:predoptcal}
Under the regression model (\ref{eqn:phyobs}) where $\zeta(\cdot)\in\HH$ and Assumption  \ref{asp:etaspace} and \ref{asp:decayrateofeigen},  for any $n\geq 1$
and any $\alpha> 0$ with probability at least $1-8e^{-\alpha^2}$,
\begin{equation*}
\begin{aligned}
& \|[\eta(\cdot,\theta^{\text{opt-pred}})+\widehat{\delta}_{n\lambda}(\cdot,\theta^{\text{opt-pred}})]- \zeta(\cdot)\|_{L_2(\Pi)}^2  \\
&\leq C_2\left[1+\alpha^{\frac{2m-d}{2m+d}}n^{-\frac{2m-d}{4m+2d}}\left(1+{\sigma}/{\|\zeta(\cdot) - \eta(\cdot,\theta^{\text{opt-pred}})\|_\HH}\right)^{-\frac{2d}{2m+d}}\right]^2\alpha^{\frac{4m}{2m+d}}\\
& \quad\cdot n^{-\frac{2m}{2m+d}}\left(\|\zeta(\cdot) - \eta(\cdot,\theta^{\text{opt-pred}})\|_\HH+\sigma\right)^2\left(1+{\sigma}/{\|\zeta(\cdot) - \eta(\cdot,\theta^{\text{opt-pred}})\|_\HH}\right)^{-\frac{2d}{2m+d}},
\end{aligned}
\end{equation*}
where $\widehat{\delta}_{n\lambda} $ is defined by (\ref{eqn:widehatdeltanlambda}) with  $\lambda = n^{-{2m}/{(2m+d)}}\{4m^{1/2}(4m-d)^{-1/2}\cdot C_\lambda^{{d}/{4m}}[\alpha A + \sigma c_\phi(1+\sqrt{2}\alpha)/\|\zeta(\cdot) - \eta(\cdot,\theta^{\text{opt-pred}})\|_\HH]\}^{{4m}/{(2m+d)}}$. Here, $C_2$ and $A$
are the same as in Theorem \ref{thm:hatfferror} that do not depend on $n,\sigma,\|\zeta-\eta(\cdot,\theta^{\text{opt-pred}})\|_\HH$.
\end{corollary}

Corollary \ref{cor:lowerbd} and \ref{cor:predoptcal} together imply that the non-asymptotic minimax optimal risk for predicting $\zeta\in\HH$ by the computer model with $\theta^{\text{opt-pred}}$ is
\begin{equation*}
\begin{aligned}
& \|[\eta(\cdot,\theta^{\text{opt-pred}})+\widetilde{\delta}(\cdot)]- \zeta\|_{L_2(\Pi)}^2  \\
& = C'  \left[1+n^{-\frac{2m-d}{4m+2d}}\left(1+\frac{\sigma}{\min_{\theta\in\Theta}\|\zeta-\eta(\cdot,\theta)\|_\HH}\right)^{-\frac{2d}{2m+d}}\right]^2\\
& \quad\cdot n^{-\frac{2m}{2m+d}}\left(\min_{\theta\in\Theta}\|\zeta-\eta(\cdot,\theta)\|_\HH+\sigma\right)^2\left(1+\frac{\sigma}{\min_{\theta\in\Theta}\|\zeta-\eta(\cdot,\theta)\|_\HH}\right)^{-\frac{2d}{2m+d}}.
\end{aligned}
\end{equation*}
Moreover,  this minimax optimal risk can be obtained by
\begin{equation}
\label{eqn:predicthatzeta}
\zeta_{n\lambda}^{\text{opt-pred}}(\cdot) \equiv \eta(\cdot, \theta^{\text{opt-pred}}) + \widehat{\delta}_{n\lambda}(\cdot, \theta^{\text{opt-pred}}),
\end{equation}
where $\widehat{\delta}_{n\lambda} $ is defined by (\ref{eqn:widehatdeltanlambda}).

\subsection{Improved prediction using the computer model}
\label{sec:improvbycompmodels}

The minimax optimal predictor (\ref{eqn:predicthatzeta}) combines the merits of parametrics (for computer model) and  nonparametrics (for the discrepancy estimator).
We compare (\ref{eqn:predicthatzeta}) with its counterpart of the minimax optimal predictor (\ref{eqn:regestforzeta}) without using the computer model.
\begin{theorem}
\label{thm:zetahhles}
Under the regression model (\ref{eqn:phyobs}) where $\zeta(\cdot)\in\HH$ and Assumption  \ref{asp:etaspace} and \ref{asp:decayrateofeigen}, if
\begin{equation}
\label{eqn:zetahhles}
\min_{\theta\in\Theta}\|\zeta(\cdot) - \eta(\cdot,\theta)\|_{\HH} < \|\zeta\|_{\HH},
\end{equation}
then $\zeta_{n\lambda}^{\text{opt-pred}}(\cdot)$ defined by  (\ref{eqn:predicthatzeta}) with the aid of the computer model achieves a smaller upper bound of the predictive mean squared  error than
$\widehat{\zeta}_{n\lambda}(\cdot)$ defined by (\ref{eqn:regestforzeta}) without using the information of the computer model.
\end{theorem}
The proof of this theorem is given in the supplementary material.  Here is a remark on the condition (\ref{eqn:zetahhles}).
The computer model $\eta(\cdot,\theta)$ is built based on some physics for the system $\zeta(\cdot)$ and Theorem \ref{thm:thetastardeltahh} shows that $\eta(\cdot, \theta^{\text{opt-pred}})$ is the best approximation to $\zeta(\cdot)$ within the family $\{\eta(\cdot,\theta),\theta\in\Theta\}$. Although the computer model is imperfect for modeling the physical system, $\eta(\cdot, \theta^{\text{opt-pred}})$  can still capture some major shape of $\zeta(\cdot)$ and consequently $\zeta(\cdot) - \eta(\cdot, \theta^{\text{opt-pred}})$ has less variation or smoother in $\HH$ than the original $\zeta(\cdot)$ does. Hence, we assume that $\|\zeta(\cdot) - \eta(\cdot,\theta^{\text{opt-pred}})\|_{\HH} < \|\zeta\|_{\HH}$, which is exactly
the condition (\ref{eqn:zetahhles}).

The predictor (\ref{eqn:predicthatzeta}) is  a parametrically-guided nonparametric predictor, where $\eta(\cdot,\theta)$ can have a parametric form.

\subsection{Comparison with existing frequentist calibration methods}
\label{sec:comparewithfrequentist}
We compare the calibration procedure in $\theta^{\text{opt-pred}}$ defined by (\ref{def:btheta0}) with two other frequentist calibration methods: the $L_2$-calibration and the least square calibration.
The $L_2$-calibration method in \cite{tuo2015efficient} is defined as follows:
\begin{equation*}
\widehat{\theta}_n^{L_2}= \underset{\theta\in\Theta}{\arg\min} \left\{\|\widehat{\zeta}_{n\lambda}(\cdot) - \eta(\cdot,\theta)\|_{L_2(\Pi)}\right\},
\end{equation*}
where $\widehat{\zeta}_{n\lambda}(\cdot)$ is defined by (\ref{eqn:regestforzeta}).
The least square calibration method in \cite{wong2017frequentist}  minimizes the model discrepancy equipped with the empirical $l_2$-norm:
\begin{equation}
\label{eqn:leastsqcal}
\widehat{\theta}_n^{l_2} = \underset{\theta\in\Theta}{\arg\min} \left\{\frac{1}{n}\sum_{i=1}^n[Y_i - \eta(X_i,\theta)]^2\right\}.
\end{equation}
In particular, \cite{wong2017frequentist}  estimates the model discrepancy after calibrating $\theta=\widehat{\theta}_n^{l_2}$:
\begin{equation*}
\widehat{\delta}_{n\lambda}(\cdot, \widehat{\theta}_n^{l_2}) = \underset{\delta(\cdot)\in\HH}{\arg\min} \left\{\frac{1}{n}\sum_{i=1}^n[Y_i - \eta(X_i,\widehat{\theta}_n^{l_2}) - \delta(X_i)] + \lambda\|\delta\|_{\HH}^2\right\},
\end{equation*}
and  a predictor for $\zeta(\cdot)$  is given by $\eta(\cdot, \widehat{\theta}_n^{l_2}) + \widehat{\delta}_{n\lambda}(\cdot, \widehat{\theta}_n^{l_2})$.

\begin{remark}
\label{prop:l2calibrationpred}
The differences between $\theta^{\text{opt-pred}}$, $\widehat{\theta}_n^{L_2}$, and $\widehat{\theta}_n^{l_2}$ are as follows.
\begin{itemize}
\item Calibration results. We have that $ \theta^{\text{opt-pred}} = \underset{\theta\in\Theta}{\arg\min}\left\{\|\zeta(\cdot) - \eta(\cdot,\theta)\|_{\HH}\right\}$ and
\begin{equation*}
\P\left\{\widehat{\theta}_n^{L_2}, \widehat{\theta}_n^{l_2} \to \theta^{L_2} \equiv\underset{\theta\in\Theta}{\arg\min}\left\{\|\zeta(\cdot) - \eta(\cdot,\theta)\|_{L_2(\Pi)}\right\} \text{ as }n\to\infty\right\}>0.
\end{equation*} Since $\widehat{\theta}_n^{L_2}$ and $\widehat{\theta}_n^{l_2}$
 converge to a different minimizer than  $\theta^{\text{opt-pred}}$,  the calibration result in $\theta^{\text{opt-pred}}$ is  different from $\widehat{\theta}_n^{L_2}$ and $\widehat{\theta}_n^{l_2}$.
\item Prediction. For a finite sample size $n$, the predictor with  $\theta^{\text{opt-pred}}$ achieves a smaller upper bound of the predictive mean squared error comparing  with the predictor with $\widehat{\theta}_n^{L_2}$ in \cite{tuo2015efficient} and the predictor with $\widehat{\theta}_n^{l_2}$ in \cite{wong2017frequentist}.
\end{itemize}
\end{remark}
We provide a proof of Remark \ref{prop:l2calibrationpred} in the supplementary material.

\section{An algorithm}
\label{sec:algorithm}
We propose an algorithm to compute the optimal calibration $\theta^{\text{opt-pred}} $ in (\ref{def:btheta0}).
From Theorem \ref{thm:thetastardeltahh},
\begin{equation}
\label{eqn:minimizehhnormdelta}
\theta^{\text{opt-pred}} =   \underset{\theta\in\Theta}{\arg\min} \left\{\|\delta(\cdot,\theta)\|^2_{\HH}\right\},
\end{equation}
where the discrepancy $\delta(\cdot,\theta)$ is subject to the constraint (\ref{eqn:discrepancymodel}).
By evaluating (\ref{eqn:discrepancymodel}) at the training data, we have
\begin{equation}
\label{eqn:deltavecxzeta}
Y_i =  \zeta(X_i) + \varepsilon_i =\eta(X_i,\theta) + \delta(X_i,\theta)+ \varepsilon_i , \ \ \forall \theta\in\Theta,i=1,\ldots,n.
\end{equation}
By using the Lagrange multiplier method for  (\ref{eqn:minimizehhnormdelta}) with the constraint (\ref{eqn:deltavecxzeta}), we need to find $\theta\in\Theta$ and $\delta(\cdot)\in\HH$ for minimizing
\begin{equation}
\label{eqn:simthetdelta}
\frac{1}{n}\sum_{i=1}^n[Y_i - \eta(X_i,\theta) - \delta(X_i)]^2 + \lambda\|\delta\|_{\HH}^2,
\end{equation}
where $\lambda> 0$ is a tuning parameter. 

The optimization problem in (\ref{eqn:simthetdelta}) can be solved iteratively as follows.
We introduce some notation. Recall that $K$ is the reproducing kernel of $(\HH,\|\cdot\|_{\HH})$. Let
$\Sigma$ be the $n\times n$ kernel matrix with $ij$th entry $K(X_i,X_j)$. Denote by  $\vec{Y} = (Y_1,\ldots,Y_n)^\top$, $\eta(\vec{X},\theta)=(\eta(X_1,\theta),\ldots,\eta(X_n,\theta))^\top$, $\delta(\vec{X},\theta)= (\delta(X_1,\theta),\ldots,\delta(X_n,\theta))^\top$, and $\zeta(\vec{X}) = (\zeta(X_1),\ldots,\zeta(X_n))^\top$.

For any fixed $\theta\in\Theta$, the minimizer $\delta(\cdot)$ of (\ref{eqn:simthetdelta}) is the same as the regularized estimator $\widehat{\delta}_{n\lambda}(\cdot,\theta)$ in (\ref{eqn:widehatdeltanlambda}). By the representer lemma  \cite{kimeldorf1971some},
\begin{equation}
\label{eqn:widehatdelnlambdafor}
\widehat{\delta}_{n\lambda}(\cdot,\theta) = \sum_{i=1}^nc_iK(X_i,\cdot),
\end{equation}
where the coefficient $c= (c_1,\ldots,c_n)^\top \in\R^n $ is given by
\begin{equation}
\label{en:cthetaformula}
c = c(\theta)= (\Sigma+n\lambda I)^{-1}[\vec{Y} - \eta(\vec{X},\theta)].
\end{equation}
The tuning parameter $\lambda$ can be selected by the
generalized cross-validation (GCV) in \cite{craven1978smoothing}. Let $A(\lambda)$ be the influence matrix satisfying $\widehat{\delta}_{n\lambda}(\vec{X},\theta) = A(\lambda) [\vec{Y} - \eta(\vec{X},\theta)]$. The GCV estimate of $\lambda$ is the minimizer of
\begin{equation}
\label{eqn:gcvlambda}
\text{GCV}(\lambda) = \frac{n^{-1}\|\vec{Y} - \eta(\vec{X},\theta) - \widehat{\delta}_{n\lambda}(\vec{X},\theta)\|^2}{[n^{-1}\text{tr}(I - A(\lambda))]^2}.
\end{equation}

GCV gives a consistent estimate of  the optimal $\lambda$ in Corollary \ref{cor:predoptcal} \cite{li1985stein, wahba1990}.
Another popular technique for selecting the tuning parameter is fivefold or tenfold cross-validation.
Since the computational load of GCV is smaller, it will be used for simulations in Section \ref{sec:simulationandareal}.

For any fixed $\delta(\cdot) = \widehat{\delta}_{n\lambda}(\cdot,\theta)$ from (\ref{eqn:widehatdelnlambdafor}), the minimizer $\theta$ of
 (\ref{eqn:simthetdelta}) is the same as
\begin{equation}
\label{eqn:nonlinearetamin}
\underset{\theta\in\Theta}{\arg\min}\left\{(\vec{Y} - \eta(\vec{X},\theta))^\top(\Sigma+n\lambda)^{-1}(\vec{Y} - \eta(\vec{X},\theta))\right\}.
\end{equation}
Since the objective function in (\ref{eqn:nonlinearetamin}) is a weighted version of the empirical $l_2$-norm $\|\vec{Y} - \eta(\vec{X},\theta)\|^2/n$,  (\ref{eqn:nonlinearetamin}) gives a different calibration result than
 the least square calibration method (\ref{eqn:leastsqcal}).

Putting the above building blocks together, our algorithm for optimizing (\ref{eqn:simthetdelta}) iterates between (\ref{eqn:widehatdelnlambdafor}) and (\ref{eqn:nonlinearetamin}).
In each iteration, (\ref{eqn:simthetdelta}) is decreased.
The algorithm can start with the calibration parameters from the least square calibration method.
Applying later iterations of the algorithm continuously improves  the initial values for prediction.
Our experience indicates that a small number of iterations is sufficient to obtain good performance of the algorithm.
In our simulation studies, we find that the objective function (\ref{eqn:simthetdelta}) decreases fast in the first iteration and becomes close to the objective function at convergence. This motivates a one-step update procedure:
\begin{itemize}
\item[1.] Initialization: Solve the least square calibration problem in (\ref{eqn:leastsqcal}) to obtain $\theta = \widehat{\theta}_n^{l_2}$.
\item[2.] Solve for $\widehat{\delta}_{n\lambda}(\cdot,\theta)$ in (\ref{eqn:widehatdelnlambdafor}), and tune $\lambda$ according to GCV. Fix $\lambda$ at the chosen value in later steps.
\item[3.] For $\widehat{\delta}_{n\lambda}(\cdot,\theta)$ obtained in step 2, solve for $\theta$ in  (\ref{eqn:nonlinearetamin}).
\item[4.] With the new $\theta$, solve for $\widehat{\delta}_{n\lambda}(\cdot,\theta)$ in (\ref{eqn:widehatdelnlambdafor}).
\end{itemize}

\subsection{Connection with Bayesian calibration}
\label{sec:comparisons}

We now link our frequentist calibration method to the Bayesian calibration method in \cite{kennedy2001bayesian}. 
\cite{kennedy2001bayesian} uses Gaussian processes \citep{sacks1989design} as priors for the computer model $\eta(\cdot,\theta)$ and the discrepancy $\delta(\cdot,\theta)$.
Recall that $K$ is the reproducing kernel of the RKHS $(\HH,\|\cdot\|_{\HH})$.
We consider the following priors
\begin{equation}
\label{eqn:bayesianmodel}
\begin{aligned}
&\zeta(x) = \eta(x,\theta)+\delta(x),   \\
&  \text{ where} \  \eta(x,\theta) = \sum_{j=1}^p\theta_jh_j(x),  \theta\sim N(0,\alpha I), \delta(x)\sim N(0,\beta K(\cdot,\cdot)).
\end{aligned}
\end{equation}
Here, $h_j(x)$ are deterministic functions and $\alpha, \beta$ are positive hyperparameters.
The following proposition builds a connection between the Bayesian calibration method and
the optimization step in (\ref{eqn:simthetdelta}) for our calibration method.

\begin{proposition}
\label{prop:equivalencewithbayesian}
Suppose that the prior for $\zeta(\cdot)$ is given by  (\ref{eqn:bayesianmodel}) and denote the posterior mean $\widehat{\zeta}_\alpha(x) = \E[\zeta(x)|Y_1,\ldots,Y_n]$.
Let the predictor corresponding to the optimal calibration be $\widehat{\zeta}_{n\lambda}^{\text{opt-pred}}(\cdot) = \eta(\cdot, \widehat{\theta}^{\text{opt-pred}}) + \widehat{\delta}_{n\lambda}(\cdot)$, which is an estimator of (\ref{eqn:predicthatzeta}) and
\begin{equation*}
(\widehat{\theta}^{\text{opt-pred}},\widehat{\delta}_{n\lambda}) =\underset{\theta\in\Theta,\delta\in\HH}{ \arg\min} \left\{\frac{1}{n}\sum_{i=1}^n[Y_i - \eta(X_i,\theta) - \delta(X_i)]^2 + \lambda\|\delta\|_{\HH}^2\right\}
\end{equation*}
with $\lambda = \sigma^2/n\beta$. Here, $\alpha$ and $\beta$ are defined in (\ref{eqn:bayesianmodel}).
Then,
$ \lim_{\alpha\to\infty}\widehat{\zeta}_\alpha(x) = \widehat{\zeta}_{n\lambda}^{\text{opt-pred}}(x)$ for any  $x\in\Omega$.
\end{proposition}

The proof of this proposition is given in the supplementary material.
The proposition sheds new lights on why the prediction of the
Bayesian calibration method works well in practice.

The Bayesian calibration method is more time consuming to compute than frequentist calibration methods such as ours.

\section{Simulation and  real  examples}
 \label{sec:simulationandareal}

We illustrate of the prediction accuracy of the proposed calibration method using several examples.
Our simulation study consists of Example \ref{exp:toypark}, \ref{exp:parkrevise}, \ref{exp:toyfallball}, where the tuning parameters for regularized estimators  are selected by the GCV.
The prediction accuracy is measured by the predictive mean squared error estimated by a Monte Carlo sample of  1000,000  test points from the same distribution as the training points.
A real data example is given in Example \ref{exp:ionchannel}.

\begin{example}
\label{exp:toypark}
Consider a physical system
$\zeta(x) = \exp(\pi x/5)\sin 2\pi x, x\in [0,1]$.
The physical data are generated by (\ref{eqn:phyobs}) with
$X_i \sim\text{Unif}([0,1]),\varepsilon_i\sim N(0,\sigma^2)$ for $i=1,\ldots, 50.$
Four different noise variances are investigated: $\sigma^2 = 0.1, 0.25, 0.5, 1$.
Suppose that the computer model is
\begin{equation*}
\begin{aligned}
\eta(x,\theta) = \zeta(x) - \sqrt{\theta^2-\theta+1}(\sin 2\pi \theta x + \cos 2\pi \theta x)\ \text{for } \  \theta\in [-1,1].
\end{aligned}
\end{equation*}
Since $\theta^2-\theta+1\geq 3/4$ for any $-1\leq \theta\leq1$, the model discrepancy between $\eta(\cdot,\theta)$ and $\zeta(\cdot)$ always exists no matter how  $\theta$ is chosen.
We use the Mat\'ern  kernel $K(x_1,x_2) = (1+|x_1-x_2|/\psi) \exp\{-|x_1-x_2|/\psi\}$, where the scale parameter $\psi$ is chosen by the five-fold cross-validation.
Figure \ref{example1} plots the squared model discrepancy  with different norms:
 $\|\zeta(\cdot) - \eta(\cdot,\theta)\|^2_{L_2(\Pi)}$  and $\|\zeta(\cdot) - \eta(\cdot,\theta)\|^2_{\HH}$.
The corresponding  minimizers  are different given as $\theta^{L_2}\approx -0.1780$ and $\theta^{\text{opt-pred}}\approx 0.3740$ (a local minimizer of $\|\zeta(\cdot) - \eta(\cdot,\theta)\|^2_{\HH}$ in $[-0.4,0]$ is $\theta^{\text{opt-pred}}\approx -0.1230$), which illustrates the first part of Theorem \ref{prop:l2calibrationpred}.
\begin{figure}[h!]
\centering
\includegraphics[width=0.9\textwidth]{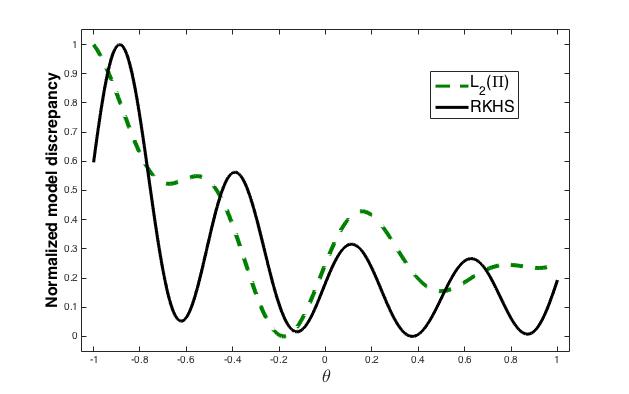}
\caption{Normalized model discrepancy equipped with $L_2(\Pi)$-norm  and RKHS-norm in Example \ref{exp:toypark}.}
\label{example1}
\end{figure}

We compare the prediction accuracy of four frequentist calibration methods:
\begin{itemize}
\item[1)] The computer model with $L_2$-calibration (abbreviated as ``\textit{No Bias Corr.}"), which is the $\eta(\cdot,\widehat{\theta}_n^{L_2})$ in Section \ref{sec:comparewithfrequentist} without the  model discrepancy correction;
\item[2)] The nonparametric predictor  (abbreviated as ``\textit{N.P.}"), which is the $\widehat{\zeta}_{n\lambda}(\cdot)$ obtained by (\ref{eqn:regestforzeta}) in Section \ref{sec:nonasymresfornonpara};
\item[3)] The predictor in \cite{wong2017frequentist} (abbreviated as ``\textit{$LS$. Cal.}"), which is the $\eta(\cdot, \widehat{\theta}_n^{l_2}) + \widehat{\delta}_{n\lambda}(\cdot, \widehat{\theta}_n^{l_2})$ in Section \ref{sec:comparewithfrequentist} based on the least square calibration;
\item[4)] Our  predictor (abbreviated as ``\textit{Opt. Cal.}"), which is the
$\zeta_{n\lambda}^{\text{opt-pred}}$ in Section \ref{sec:apptooptcalibration} based on the proposed calibration and computed by the algorithm in Section \ref{sec:algorithm}.
 \end{itemize}
For each chosen noise variance, we replicate the data generation, calibration and prediction procedures 1,000 times and average the results for each method across the replicates.
The resulting average predictive mean squared errors and its associated standard errors are given in Table \ref{table:tuowuexample1}.
The computer model with $L_2$-calibration (i.e., ``\textit{No Bias Corr.}") gives the largest predictive mean squared error, which shows that an estimator for the model discrepancy is necessary. ``\textit{No Bias Corr.}" has the smallest standard error which is not surprising because  the parametric estimation of $\theta$ here has a faster rate of convergences than nonparametric estimations required for other three methods.
Table \ref{table:tuowuexample1} indicates that our method ``\textit{Opt. Cal.}" and the predictor in \cite{wong2017frequentist}  ``\textit{$LS$. Cal.}" outperform the nonparametric predictor ``\textit{N.P.}". This  advantage illustrates the improved predictions by computer models as indicated in Theorem \ref{thm:zetahhles}.  Furthermore, ``\textit{Opt. Cal.}"  gives smaller prediction errors than  ``\textit{$LS$. Cal.}", which agrees with the second part of Theorem \ref{prop:l2calibrationpred}.
Overall, for a finite sample size $n=50$ , the proposed method ``\textit{Opt. Cal.}" outperforms the other frequentist predictors in the settings studied.

\begin{table}[h!]
\centering
 \caption{Comparison of predictive mean squared errors for Example \ref{exp:toypark}.  PMSE = predictive mean squared error, SE = standard error}
 \begin{tabular}{l  c c  c c}
 \hline
  \rule{0pt}{\normalbaselineskip}
       & \multicolumn{2}{c}{$\sigma^2=0.1$} & \multicolumn{2}{c}{$\sigma^2=0.25$}   \\
     & Average of PMSE & SE of PMSE  & Average of PMSE & SE of PMSE  \\   [0.5ex]
 \hline
 \rule{0pt}{\normalbaselineskip}
 No Bias Corr. & $0.3492$ & $0.0174$ & $0.3601$ & $0.0270$ \\  [0.5ex] \cline{2-5}
 \rule{0pt}{\normalbaselineskip}
 N.P. & $0.1701$ & $0.0594$ & $0.2327$ & $0.0680$ \\  [0.5ex] \cline{2-5}
 \rule{0pt}{\normalbaselineskip}
 LS. Cal. & $0.1152$ & $0.0604$ & $0.1693$ & $0.0580$ \\  [0.5ex] \cline{2-5}
     \rule{0pt}{\normalbaselineskip}
Opt. Cal.   & \boldsymbol{$0.0922$} & $0.0515$ & $\boldsymbol{0.1434}$ & $0.0552$ \\ [0.5ex] \cline{2-5}
 \hline
  \hline
  \rule{0pt}{\normalbaselineskip}
       & \multicolumn{2}{c}{$\sigma^2=0.5$} & \multicolumn{2}{c}{$\sigma^2=1$}   \\
     & Average of PMSE & SE of PMSE  & Average of PMSE & SE of PMSE  \\   [0.5ex]
 \hline
 \rule{0pt}{\normalbaselineskip}
 No Bias Corr. & $0.3744$ & $0.0407$ & $0.3985$ & $0.0666$ \\  [0.5ex] \cline{2-5}
 \rule{0pt}{\normalbaselineskip}
 N.P. & $0.2810$ & $0.0930$ & $0.3417$ & $0.1367$ \\  [0.5ex] \cline{2-5}
 \rule{0pt}{\normalbaselineskip}
 LS. Cal. & $0.1924$ & $0.0683$ & $0.2074$ & $0.0853$ \\  [0.5ex] \cline{2-5}
     \rule{0pt}{\normalbaselineskip}
Opt. Cal.   & \boldsymbol{$0.1684$} & $0.0656$ & $\boldsymbol{0.1838}$ & $0.0788$ \\ [0.5ex] \cline{2-5}
 \hline
 \end{tabular}
 \label{table:tuowuexample1}
\end{table}

\end{example}

\begin{example}
\label{exp:parkrevise}
Consider a two-dimensional physical system
\begin{equation*}
\begin{aligned}
\zeta(x_1,x_2) & = \frac{2}{3}\exp(x_1+0.2) - x_2\sin 0.4 + 0.4 \\
&  \quad+ \exp(-x_1)\left(x_1+\frac{1}{2}\right)\left(x_2^2+x_2+1\right),\quad (x_1,x_2)\in [0,1]^2.
\end{aligned}
\end{equation*}
Suppose that the computer model is
\begin{equation*}
\eta(x_1,x_2;\theta_1,\theta_2)  = \frac{2}{3}\exp(x_1+\theta_1) - x_2\sin \theta_2 +\theta_2,\quad (\theta_1,\theta_2)\in[0,1]^2.
\end{equation*}
The model discrepancy exists between $\eta(\cdot;\theta_1,\theta_2)$ and $\zeta(\cdot)$.
Assume the physical data are generated by (\ref{eqn:phyobs}) with a uniform design on $[0,1]^2$ and $n=50$.
We consider four levels of $\sigma^2=0.03, 0.05, 0.07, 0.1$, and use the Mat\'ern kernel $K(x_1,x_2) = (1+\|x_1-x_2\|/\psi) \exp\{-\|x_1-x_2\|/\psi\}$ with $\psi$ chosen by the five-fold cross-validation.
For each level of $\sigma^2$, we replicate the data generation, calibration and prediction procedures 1,000 times for all the methods and average the results for each method across the replicates.
Table \ref{table:parkrevise} summarizes the results, where the abbreviations of the methods are the same as those in  Example \ref{exp:toypark}.
Once again, the proposed method ``\textit{Opt. Cal.}" outperforms the other three frequentist calibration methods in terms of the predictive mean squared error.

\begin{table}[h!]
\centering
 \caption{Comparison of predictive mean squared errors for Example \ref{exp:parkrevise}.  PMSE = predictive mean squared error, SE = standard error}
 \begin{tabular}{l  c c  c c}
 \hline
  \rule{0pt}{\normalbaselineskip}
       & \multicolumn{2}{c}{$\sigma^2=0.03$} & \multicolumn{2}{c}{$\sigma^2=0.05$}   \\
     & Average of PMSE & SE of PMSE  & Average of PMSE & SE of PMSE  \\   [0.5ex]
 \hline
 \rule{0pt}{\normalbaselineskip}
  No Bias Corr. & $0.1690$ & $0.0027$ & $0.1691$ & $0.0027$ \\  [0.5ex] \cline{2-5}
 \rule{0pt}{\normalbaselineskip}
 N.P. & $0.1155$ & $0.0512$ & $0.1823$ & $0.0850$ \\  [0.5ex] \cline{2-5}
 \rule{0pt}{\normalbaselineskip}
 LS. Cal. & $0.0611$ & $0.0198$ & $0.0743$ & $0.0212$ \\  [0.5ex] \cline{2-5}
     \rule{0pt}{\normalbaselineskip}
Opt. Cal.   & \boldsymbol{$0.0564$} & $0.0187$ & \boldsymbol{$0.0691$} & $0.0205$ \\ [0.5ex] \cline{2-5}
 \hline
  \hline
  \rule{0pt}{\normalbaselineskip}
       & \multicolumn{2}{c}{$\sigma^2=0.07$} & \multicolumn{2}{c}{$\sigma^2=0.1$}   \\
     & Average of PMSE & SE of PMSE  & Average of PMSE & SE of PMSE  \\   [0.5ex]
 \hline
 \rule{0pt}{\normalbaselineskip}
  No Bias Corr. & $0.1692$ & $0.0028$ & $0.1694$ & $0.0028$ \\  [0.5ex] \cline{2-5}
 \rule{0pt}{\normalbaselineskip}
 N.P. & $0.2425$ & $0.1159$ & $0.3327$ & $0.1453$ \\  [0.5ex] \cline{2-5}
 \rule{0pt}{\normalbaselineskip}
 LS. Cal. & $0.0825$ & $0.0224$ & $0.0906$ & $0.0235$ \\  [0.5ex] \cline{2-5}
     \rule{0pt}{\normalbaselineskip}
Opt. Cal.   & \boldsymbol{$0.0776$} & $0.0220$ & \boldsymbol{$0.0863$} & $0.0234$ \\ [0.5ex] \cline{2-5}
 \hline
 \end{tabular}
 \label{table:parkrevise}
\end{table}

\end{example}

\begin{example}
\label{exp:toyfallball}
We now compare the proposed calibration method with some Bayesian calibration method.
Consider a falling ball example in \cite{plumlee2017bayesian} where the physical system is
\begin{equation*}
\zeta(x) = 8+ \frac{5}{2}\log\left(\frac{50}{49} - \frac{50}{49}\tanh\left(\tanh^{-1}(\sqrt{0.02}) + \sqrt{2}x\right)^2\right), \ x\in[0,1]
\end{equation*}
and the computer model derived from Newton's second law is $\eta(x; v_0,g) = 8 + v_0x-gx^2/2$.
Here, calibration parameters $(v_0,g)$ are the vertical velocity and the acceleration rate, respectively.
The model discrepancy exists between $\zeta(\cdot)$ and $\eta(\cdot;v_0,g)$.
Suppose that the physical data are generated by (\ref{eqn:phyobs}) with a uniform design on $[0,1]$ and $n=30$.
We compare the proposed method ``\textit{Opt. Cal.}" with two Bayesian methods in terms of prediction accuracy:
\begin{itemize}
\item[1)] The Bayesian method in Kennedy and O'Hagan \cite{kennedy2001bayesian} (abbreviated as ``\textit{KO}");
\item[2)] The Bayesian predictor
using an orthogonal Gaussian process in $L_2$-norm as the prior (abbreviated as ``\textit{OGP}") proposed by  \cite{plumlee2017bayesian}.
\end{itemize}
The Mat\'ern kernel $K(x_1,x_2) = (1+|x_1-x_2|/\psi) \exp\{-|x_1-x_2|/\psi\}$ with parameter $\psi=1$ is used as the reproducing kernel for ``\textit{Opt. Cal.}" and the prior covariance function for both ``\textit{KO}" and ``\textit{OGP}".
Four levels of $\sigma^2=0.0025, 0.01, 0.0625, 0.25$ are considered.
For each level of $\sigma^2$, we replicate the data generation, calibration and prediction procedures 1,000 times.
Table \ref{table:fallingball} summarizes the prediction results.
Here ``\textit{KO}" has large prediction errors and large
posterior variances compared with ``\textit{OGP}" and ``\textit{Opt. Cal.}".
``\textit{OGP}" provides stable and accurate predictions and our method ``\textit{Opt. Cal.}" gives even smaller prediction errors for some $\sigma^2$ levels.

\begin{table}[h!]
\centering
 \caption{Comparison of predictive mean squared errors for Example \ref{exp:toyfallball}.  PMSE = predictive mean squared error, SE = standard error}
 \begin{tabular}{l  c c  c c}
 \hline
  \rule{0pt}{\normalbaselineskip}
      & \multicolumn{2}{c}{$\sigma^2=0.1$} & \multicolumn{2}{c}{$\sigma^2=0.2$}   \\
     & Average of PMSE & SE of PMSE & Average of PMSE & SE of PMSE   \\   [0.5ex]
 \hline
    \rule{0pt}{\normalbaselineskip}
  KO  & $0.5413$ & $2.3944$ & $5.8596$ & $29.0264$ \\  [0.5ex] \cline{2-5}
    \rule{0pt}{\normalbaselineskip}
OGP  & $0.0147$ & $0.0132$ & $0.0230$ & $0.0249$ \\  [0.5ex] \cline{2-5}
     \rule{0pt}{\normalbaselineskip}
Opt. Cal.  & \boldsymbol{$0.0091$} & ${0.0084}$ & \boldsymbol{$0.0166$} & {$0.0168$} \\ [0.5ex] \cline{2-5}
 \hline \hline
  \rule{0pt}{\normalbaselineskip}
        & \multicolumn{2}{c}{$\sigma^2=0.4$} & \multicolumn{2}{c}{$\sigma^2=0.8$}   \\
     & Average of PMSE & SE of PMSE & Average of PMSE & SE of PMSE   \\   [0.5ex]
 \hline
    \rule{0pt}{\normalbaselineskip}
  KO  & $1.4644$ & $4.3802$ & $18.5433$ & $97.3962$\\  [0.5ex] \cline{2-5}
    \rule{0pt}{\normalbaselineskip}
OGP  & $0.0500$ & $0.0478$ & $0.0952$ &  $0.0995$\\  [0.5ex] \cline{2-5}
     \rule{0pt}{\normalbaselineskip}
Opt. Cal.  & $\boldsymbol{0.0318}$ & {$0.0338$} &  \boldsymbol{$0.0620$} &  {$0.0677$} \\ [0.5ex] \cline{2-5}
 \hline
 \end{tabular}
 \label{table:fallingball}
\end{table}

\end{example}

\begin{example}[Real data example]
\label{exp:ionchannel}

We analyze a real dataset from a single voltage clamp experiment  on sodium ion channels of cardiac cell membranes.
This dataset consists of $19$ outputs and  is from \cite{plumlee2017bayesian}.
The response variable is  the normalized current for maintaining a fixed membrane potential of $-35$mV and the input variable is the logarithm of time.
Suppose the computer model for this experiment is
the Markov model for sodium ion channels given by
$\eta(x,\theta) = e_1^\top\exp(\exp(x)A(\theta))e_4,$
where $\theta = (\theta_1,\theta_2,\theta_3)^\top\in\R^3$, $e_1 = (1,0,0,0)^\top$,  $e_4 = (0,0,0,1)^\top$, and
\begin{equation*}
A(\theta) = \left( \begin{array}{cccc}
-\theta_2 - \theta_3 & \theta_1 & 0 & 0 \\
\theta_2 & - \theta_1 - \theta_2 &  \theta_1 & 0 \\
0 &  \theta_2 & - \theta_1 - \theta_2   & \theta_1 \\
0 & 0 & \theta_2 & -\theta_1 \end{array} \right)
\end{equation*}
For this example, we compare the frequentist methods in Example \ref{exp:toypark} and \ref{exp:parkrevise}, and Bayesian methods in Example \ref{exp:toyfallball}.
The Mat\'ern kernel $K(x_1,x_2) = (1+|x_1-x_2|/\psi) \exp\{-|x_1-x_2|/\psi\}$ with $\psi=1$ is used for all methods
and  the Metropolis-Hastings  algorithm is applied to sample from the posterior of $\theta$ for Bayesian methods.
In each experiment, we perform five-fold cross-validation where
the data is randomly partitioned into five roughly equal-sized parts: four parts are  for training and the rest part is  for testing.
The cross-validation process is  repeated five times, with each of the five parts is used  once for testing.
Then, the  five predictive mean squared errors are
 averaged to give a single predictive mean squared error.
We replicate the data generation, calibration and prediction procedure  100 times and average the results.

Table \ref{table:ionexamplecomparisons} summarizes the prediction results, with the abbreviations of the methods given in Examples \ref{exp:toypark} and \ref{exp:toyfallball}.
``\textit{No Bias Corr.}" gives the largest predictive mean squared error among the four frequentist methods,  indicating the existence of model discrepancy.
Here all  the frequentist methods outperform the two Bayesian methods.
``\textit{No Bias Corr.}" outperforms ``\textit{LS. Cal.}". This indicates that if the calibration parameter is not chosen well, the use of  computer model does not improve  prediction.  Overall,
the proposed method  `\textit{Opt. Cal.}" gives the smallest prediction error among all the methods applied to this example.

\begin{table}[h!]
\centering
 \caption{Comparison of predictive mean squared errors for Example \ref{exp:ionchannel}.  PMSE = predictive mean squared error, SE = standard error}
 \begin{tabular}{l  c c }
 \hline
  \rule{0pt}{\normalbaselineskip}
     & Average of PMSE & SE of PMSE  \\   [0.5ex]
 \hline
 \rule{0pt}{\normalbaselineskip}
   KO  & $0.0045$ & $0.0131$ \\  [0.5ex] \cline{2-3}
    \rule{0pt}{\normalbaselineskip}
OGP  & $9.4387\times 10^{-4}$ & $0.0022$ \\  [0.5ex] \cline{2-3}
     \rule{0pt}{\normalbaselineskip}
  No Bias Corr. & $ 4.2823\times 10^{-4}$ & $ 6.2333\times 10^{-4}$ \\  [0.5ex] \cline{2-3}
 \rule{0pt}{\normalbaselineskip}
 N.P. & $2.5916\times 10^{-4}$ & $2.2522\times 10^{-4}$ \\  [0.5ex] \cline{2-3}
 \rule{0pt}{\normalbaselineskip}
 LS. Cal. & $3.0521\times 10^{-4}$ & $6.3797\times 10^{-4}$ \\  [0.5ex] \cline{2-3}
   \rule{0pt}{\normalbaselineskip}
Opt. Cal.   & $\boldsymbol{1.6323\times 10^{-4}}$ & $2.1335\times 10^{-4}$ \\
 \hline
 \end{tabular}
 \label{table:ionexamplecomparisons}
\end{table}
\end{example}

\section{Concluding remarks and discussions}\label{sec:concludingremarks}

We have proposed a new look at the model calibration issue in computer models. This viewpoint simultaneously considers two facts regarding how computer models are used in practice: computer models are inadequate for physical systems, regardless how the calibration parameters are tuned; and only a finite number of data points are available from the physical experiment associated with a computer model. We establish a non-asymptotic minimax theory and derive an optimal prediction-oriented calibration method.  Through several examples, the proposed calibration method has some advantages in prediction when compared with some existing calibration methods.
We have developed  an algorithm  to carry out the proposed calibration method and built a link between our method and the Bayesian calibration method.  
Beyond calibration of computer models, our method can be applied to calibrate unknown parameters for general misspecified models in statistics and engineering.
In many applications, bounded linear functional information such as derivative data are observed or can be easily calculated together with the function observations. It would be interesting to include all these data in our proposed calibration method.



\appendix

\begin{supplement}\label{supplement}
\stitle{Supplement to ``Another look at Statistical Calibration: Non-Asymptotic Theory and Prediction-Oriented  Optimality".}
\slink[doi]{}
\sdatatype{.pdf}
\sdescription{The supplementary materials contain proofs of technical results in this paper.}
\end{supplement}


\end{document}


\begin{frontmatter}
\title{Supplement to ``Another look at Statistical Calibration: Non-Asymptotic Theory and Prediction-Oriented  Optimality"}
\runtitle{Supplement}
\begin{aug}
\author{\fnms{Xiaowu} \snm{Dai}\thanksref{t1, t2}
\ead[label=e1]{xdai26@wisc.edu}}
\and
\author{\fnms{Peter} \snm{Chien}\thanksref{t2}
\ead[label=e2]{	peter.chien@wisc.edu}}

\thankstext{t1}{Supported in part by NSF Grant DMS-1308877.}
\thankstext{t2}{Supported in part by NSF Grant DMS-1564376.}
\runauthor{X. Dai and P. Chien}

\affiliation{University of Wisconsin-Madison}

\address{Department of Statistics\\
University of Wisconsin-Madison\\
1300 University Avenue\\
Madison, Wisconsin 53706\\
USA\\
\printead{e1}\\
\phantom{E-mail:\ }\printead*{e2}}
\end{aug}

\end{frontmatter}

\appendix

\section{Proofs of technical results}\label{app}

This appendix consists of three parts. In Section \ref{sec:proofsec3}, we gives proofs for main results   in Section \ref{sec:finitesampleproperies}. In Section \ref{sec:proofsec4}, we prove the result in Section \ref{sec:algorithm}.
In Section \ref{sec:keylemma}, we present the key lemma used  in Section \ref{sec:proofsec3}.


\subsection{Proofs for Section \ref{sec:finitesampleproperies}}
\label{sec:proofsec3}

\subsubsection{Lower bound result: Theorem \ref{thm:lowerbdnonasymnonpara}}
Let $N$ be a natural number to be defined later and $b=\{b_{\nu}: \nu=1,\ldots,N\}\in\{0,1\}^N$ be a length-$N$ binary sequence. Recall that $\lambda_\nu$s and $\phi_\nu(\cdot)$s are the eigenvalues and eigenfunctions of $K(\cdot,\cdot)$, respectively, and they satisfy Assumption \ref{asp:decayrateofeigen}.
We define a set of functions $\zeta_b(\cdot)$ indexed by $b$ as follows:
\begin{equation*}
\begin{aligned}
 \zeta_b(x)  = & N^{-1/2}[1+N^{-\frac{2m-d}{2d}}\left(1+{\sigma}/{\|\zeta\|_\HH}\right)^{-\frac{2d}{2m+d}}]\\
 & \quad \cdot (\|\zeta\|_\HH+\sigma)(1+\sigma/\|\zeta\|_\HH)^{-d/(2m+d)}\sum_{\nu=1}^Nb_{\nu}\lambda_{\nu+N}^{1/2}\phi_{\nu+N}(x).
\end{aligned}
\end{equation*}
It is clear that
\begin{equation*}
\begin{aligned}
\|\zeta_b\|_\HH^2  & = N^{-1}[1+N^{-\frac{2m-d}{2d}}\left(1+{\sigma}/{\|\zeta\|_\HH}\right)^{-\frac{2d}{2m+d}}]^2\\
& \quad\quad\cdot(\|\zeta\|_\HH+\sigma)^2(1+\sigma/\|\zeta\|_\HH)^{-2d/(2m+d)}\sum_{\nu=1}^{N}b_{\nu}^2\\
& \leq [1+N^{-\frac{2m-d}{2d}}\left(1+{\sigma}/{\|\zeta\|_\HH}\right)^{-\frac{2d}{2m+d}}]^2\\
& \quad\quad\cdot(\|\zeta\|_\HH+\sigma)^2(1+\sigma/\|\zeta\|_\HH)^{-2d/(2m+d)}<\infty,
\end{aligned}
\end{equation*}
which also implies $\zeta_b(\cdot)\in\HH$.
By the Varshamov-Gilbert bound, there exists a collection of binary sequences $\{b^{(1)},\ldots,b^{(M)}\}\subset\{0,1\}^{N}$ such that $M\geq e^{N/8}$ with the pairwise Hamming distance satisfying
\begin{equation*}
H(b^{(l)},b^{(q)})\geq  N/8,\quad \forall 1\leq l<q\leq M.
\end{equation*}
From Assumption \ref{asp:decayrateofeigen},   we have that for any $b^{(l)},b^{(q)}\in\{0,1\}^{N}$,
\begin{equation*}
\begin{aligned}
& \|\zeta_{b^{(l)}}-\zeta_{b^{(q)}}\|_{L_2(\Pi)}^2 \\
&  \geq c_\lambda N^{-1}[1+N^{-\frac{2m-d}{2d}}\left(1+{\sigma}/{\|\zeta\|_\HH}\right)^{-\frac{2d}{2m+d}}]^2\\
& \quad\cdot(\|\zeta\|_\HH+\sigma)^2(1+\sigma/\|\zeta\|_\HH)^{-2d/(2m+d)} (2N)^{-2m/d}\sum_{\nu=1}^{N}\left[b^{(l)}_{\bnu} - b^{(q)}_{{\bnu}}\right]^2\\
& \geq c_\lambda N^{-1}[1+N^{-\frac{2m-d}{2d}}\left(1+{\sigma}/{\|\zeta\|_\HH}\right)^{-\frac{2d}{2m+d}}]^2\\
& \quad\cdot(\|\zeta\|_\HH+\sigma)^2(1+\sigma/\|\zeta\|_\HH)^{-2d/(2m+d)} (2N)^{-2m/d}N/8 \\
& = c_\lambda [1+N^{-\frac{2m-d}{2d}}\left(1+{\sigma}/{\|\zeta\|_\HH}\right)^{-\frac{2d}{2m+d}}]^2\\
& \quad\cdot 2^{-3-2m/d} N^{-2m/d} (\|\zeta\|_\HH+\sigma)^2(1+\sigma/\|\zeta\|_\HH)^{-2d/(2m+d)}.
\end{aligned}
\end{equation*}
On the other hand, again from Assumption \ref{asp:decayrateofeigen}, we have that for any $b^{(l)}\in\{b^{(1)},\ldots,b^{(M)}\}$,
\begin{equation*}
\begin{aligned}
&\|\zeta_{b^{(l)}}\|_{L_2(\Pi)}^2 \\
&  \leq C_\lambda  N^{-1}[1+N^{-\frac{2m-d}{2d}}\left(1+{\sigma}/{\|\zeta\|_\HH}\right)^{-\frac{2d}{2m+d}}]^2\\
& \quad\cdot(\|\zeta\|_\HH+\sigma)^2(1+\sigma/\|\zeta\|_\HH)^{-2d/(2m+d)} N^{-2m/d}\sum_{\nu=1}^{N}\left[b^{(l)}_{\vec{\bnu}} \right]^2\\
&\leq C_\lambda N^{-2m/d}[1+N^{-\frac{2m-d}{2d}}\left(1+{\sigma}/{\|\zeta\|_\HH}\right)^{-\frac{2d}{2m+d}}]^2\\
& \quad\cdot(\|\zeta\|_\HH+\sigma)^2(1+\sigma/\|\zeta\|_\HH)^{-2d/(2m+d)}.
\end{aligned}
\end{equation*}

It is known that the lower bound can be reduced to the error probability in a multi-way hypothesis test (see, e.g., \cite{cover2012elements}). Let $\Theta$ be a random variable uniformly distributed on $\{1,\ldots,M\}$, then
\begin{equation}
\label{eqn:inftildzetainhhl2pi}
\begin{aligned}
& \inf_{\widetilde{\zeta}}\sup_{\zeta\in\HH}\P\left\{\|\widetilde{\zeta} - \zeta\|_{L_2(\Pi)}^2\geq \frac{1}{4}\min_{b^{(l)}\neq b^{(q)}}\|\zeta_{b^{(l)}} - \zeta_{b^{(q)}}\|^2_{L_2(\Pi)}\right\}\\
&\geq \inf_{\widehat{\Theta}}\P\{\widehat{\Theta}\neq \Theta\}\\
&= \inf_{\widehat{\Theta}}\E_{X_1,\ldots,X_n} \P\left\{\widehat{\Theta}\neq \Theta|X_1,\ldots,X_n \right\},
\end{aligned}
\end{equation}
where the infimum on the right-hand side is with respect to all  measurable functions of the data. By Fano's lemma, we have
\begin{equation}
\label{eqn:fanolemhatthetap}
\begin{split}
& \P\left\{\widehat{\Theta}\neq \Theta|X_1,\ldots,X_n \right\}\geq 1-\frac{\mathbbm{1}_{X_1,\ldots,X_n}(Y_1,\ldots,Y_n;\Theta)+\log 2}{\log M},
\end{split}
\end{equation}
where $\mathbbm{1}_{X_1,\ldots,X_n}(Y_1,\ldots,Y_n)$ is
conditioning on $\{X_1,\ldots,X_n\}$ the
 mutual information between $\Theta$ and $\{Y_1,\ldots,Y_n\}$. Let  $\mathcal{K}(\cdot|\cdot)$ be the Kullback-Leibler distance and  $\mathbf{P}_{\zeta}$ be the conditional distribution of $Y_i$s given $\{X_1,\ldots,X_n\}$. Thus,
\begin{equation*}
\begin{aligned}
& \E_{X_1,\ldots,X_n}\left[\mathbbm{1}_{X_1,\ldots,X_n}\left(Y_1,\ldots,Y_n;\Theta\right)\right]\\
&  \leq \binom M2^{-1} \sum_{b^{(l)}\neq b^{(q)}} \E_{X_1,\ldots,X_n} \mathcal{K}\left(\mathbf{P}_{\zeta_{b^{(l)}}}| \mathbf{P}_{\zeta_{b^{(q)}}}\right)\\
& \leq \frac{n}{2}\binom M2^{-1} \sum_{b^{(l)}\neq b^{(q)}} \E_{X_1,\ldots,X_n}\left[\frac{1}{n\sigma^2}\sum_{i=1}^n\left(\zeta_{b^{(l)}}(X_i) - \zeta_{b^{(q)}}(X_i)\right)^2\right]\\
& \leq \frac{n}{2\sigma^2}\binom M2^{-1} \sum_{b^{(l)}\neq b^{(q)}} \|\zeta_{b^{(l)}} - \zeta_{b^{(q)}}\|_{L_2(\Pi)}^2 \\
&  \leq \frac{n}{2\sigma^2} \max_{b^{(l)}\neq b^{(q)}} \|\zeta_{b^{(l)}} - \zeta_{b^{(q)}}\|_{L_2(\Pi)}^2 \\
&  \leq \frac{2n}{\sigma^2}\max_{b^{(l)}\in\{b^{(1)},\ldots,b^{(M)}\}} \|\zeta_{b^{(l)}}\|_{L_2(\Pi)}^2   \\
&\leq \frac{2n}{\sigma^2}C_\lambda N^{-2m/d}[1+N^{-\frac{2m-d}{2d}}\left(1+{\sigma}/{\|\zeta\|_\HH}\right)^{-\frac{2d}{2m+d}}]^2\\
& \quad\cdot(\|\zeta\|_\HH+\sigma)^2(1+\sigma/\|\zeta\|_\HH)^{-2d/(2m+d)}.
\end{aligned}
\end{equation*}
Therefore, (\ref{eqn:inftildzetainhhl2pi}) and (\ref{eqn:fanolemhatthetap}) imply that
\begin{equation*}
\begin{aligned}
& \inf_{\widetilde{\zeta}}\sup_{\zeta\in\HH}\P\left\{\|\widetilde{\zeta} - \zeta\|_{L_2(\Pi)}^2\geq c_\lambda  [1+N^{-\frac{2m-d}{2d}}\left(1+{\sigma}/{\|\zeta\|_\HH}\right)^{-\frac{2d}{2m+d}}]^2\right.\\
&\quad\quad\quad\quad\quad\quad\quad\left.\cdot2^{-5-2m/d}N^{-2m/d}(\|\zeta\|_\HH+\sigma)^2(1+\sigma/\|\zeta\|_\HH)^{-2d/(2m+d)}\right\} \\
&  \geq 1-\frac{1}{\log M}\left[\E\mathbbm{1}_{X_1,\ldots,X_n}(Y_1,\ldots,Y_n;\Theta)+\log 2\right]\\
& \geq 1-\frac{16C_\lambda n(\|\zeta\|_\HH+\sigma)^2[1+N^{-\frac{2m-d}{2d}}\left(1+{\sigma}/{\|\zeta\|_\HH}\right)^{-\frac{2d}{2m+d}}]^2 }{(1+\sigma/\|\zeta\|_\HH)^{2d/(2m+d)}N^{1+2m/d}\sigma^2} - \frac{8\log 2}{N}.
\end{aligned}
\end{equation*}
Let $N=cn^{d/(2m+d)}$ for some large $c>0$, we have
\begin{equation*}
\begin{aligned}
& \inf_{\widetilde{\zeta}} \sup_{\zeta\in\HH}\P\left\{\|\widetilde{\zeta} - \zeta\|_{L_2(\Pi)}^2 \geq  c_\lambda [1+n^{-\frac{2m-d}{4m+2d}}\left(1+{\sigma}/{\|\zeta\|_\HH}\right)^{-\frac{2d}{2m+d}}]^2\cdot\right.\\
&\quad \ \ \left. c^{-2m/d}2^{-5-2m/d}  n^{-2m/(2m+d)} (\|\zeta\|_\HH+\sigma)^2(1+\sigma/\|\zeta\|_\HH)^{-2d/(2m+d)}\right\}\\
&\geq 1-\frac{16C_\lambda(\|\zeta\|_\HH+\sigma)^2[1+n^{-\frac{2m-d}{4m+2d}}\left(1+{\sigma}/{\|\zeta\|_\HH}\right)^{-\frac{2d}{2m+d}}]^2}{c^{4m/d}\sigma^2(1+\sigma/\|\zeta\|_\HH)^{2d/(2m+d)}}-\frac{8\log 2}{cn^{d/(2m+d)}}\\
& >0.
\end{aligned}
\end{equation*}
This completes the proof.

\subsubsection{Upper bound result: Theorem \ref{thm:hatfferror}}
We define a new norm $\|\cdot\|$ in $\HH$ by
\begin{equation*}
\|g\|^2 = \|g\|_{L_2(\Pi)}^2 +\|g\|^2_{\HH},\quad\forall g\in\HH.
\end{equation*}
Note that $\|\cdot\|$ is a norm because that $\|\cdot\|^2$ defined above is a quadratic form and it equals  to  zero if and only if $g = 0$.
Since the density function of $\Pi$ is bounded away from zero and infinity, there exists some constant $c>0$ such that $\|g\|_{L_2(\Pi)}^2\leq c\|g\|_{\HH}^2$. Thus, $\|g\|^2\leq (c+1)\|g\|_{\HH}^2$. This together with the fact $\|g\|_{\HH}^2\leq \|g\|^2$ imply that  $\|\cdot\|$ and $\|\cdot\|_{\HH}$ are equivalent. In particular,  $\|g\|<\infty$ if and only if $\|g\|_{\HH}<\infty$.
Let $\langle\cdot,\cdot\rangle$ be the inner product associated with $\|\cdot\|$, which can be constructed as follows:
\begin{equation*}
\langle g_1, g_2\rangle = \frac{1}{4} (\|g_1+g_2\|^2 - \|g_1-g_2\|^2), \quad \forall g_1,g_2\in\HH.
\end{equation*}
Denote by $R(\cdot,\cdot)$ the reproducing kernel associated with $(\HH,\|\cdot\|)$. We have the following eigenvalue decomposition:
\begin{equation*}
R(x,x') = \sum_{\nu\geq 1}(1+\lambda^{-1}_\nu)^{-1}\phi_\nu(x)\phi_\nu(x').
\end{equation*}
Let $g_\nu = \langle g,\phi_\nu\rangle_{L_2(\Pi)}$ for any $g\in\HH$, then
\begin{equation*}
\|g\|^2 = \sum_{\nu=1}^\infty(1+\lambda^{-1}_\nu)g_\nu^2, \quad \|g\|_{L_2(\Pi)}^2 = \sum_{\nu=1}^\infty g_\nu^2,
\quad \|g\|_{\HH}^2  = \sum_{\nu=1}^\infty \lambda^{-1}_\nu g_\nu^2.
\end{equation*}
Now, we define a norm $\|\cdot\|_a$ for any $0\leq a\leq 1$ by
\begin{equation*}
\|g\|_a^2 = \sum_{\nu=1}^\infty(1+\lambda_\nu^{-a})g_\nu^2.
\end{equation*}
It is clear that $\|g\|_0 = \sqrt{2}\|g\|_{L_2(\Pi)}$ and $\|g\|_1 = \|g\|$. Let $\langle\cdot,\cdot\rangle_a$ be the inner product associated with $\|\cdot\|_a$ for any $0\leq a\leq 1$.

We write
\begin{equation*}
l_{n}(g) = \frac{1}{2n}\sum_{i=1}^n[Y_i-g(X_i)]^2,
\end{equation*}
and  $l_{n\lambda}(g) = l_n(g) + \frac{1}{2}\lambda\|g\|_{\HH}^2.$
Then, $\widehat{\zeta}_{n\lambda} = \arg\min_{g\in \HH}l_{n\lambda}(g)$.
Denote by $l_\infty(g) = \E[l_n(g)]$, then $l_\infty(g) = \frac{1}{2}(\sigma^2+\|\zeta-g\|^2_{L_2(\Pi)})$. Write that
\begin{equation*}
\bar{\zeta}_{\infty\lambda} = \underset{g\in\HH}{\arg\min}\ l_{\infty\lambda}(g), \text{ where } l_{\infty\lambda}(g) = l_\infty(g) + \frac{1}{2}\lambda\|g\|_\HH^2.
\end{equation*}
Now we can decompose the estimation error of $\widehat{\zeta}_{n\lambda}$ as follows:
\begin{equation*}
\widehat{\zeta}_{n\lambda} - \zeta = (\widehat{\zeta}_{n\lambda} - \bar{\zeta}_{\infty\lambda})  + (\bar{\zeta}_{\infty\lambda} - \zeta).
\end{equation*}
Here, the two terms on the right-hand side are referred to as the stochastic error and the deterministic error, respectively. We study these two terms separately in the following.

\paragraph{Step 1: Deterministic error}
Denote by $\zeta_\nu = \langle \zeta,\phi_\nu\rangle_{L_2(\Pi)}$. Then, $\zeta(\cdot) = \sum_{\nu=1}^\infty \zeta_\nu\phi_\nu(\cdot)$. It is clear that
\begin{equation*}
\bar{\zeta}_{\infty\lambda} = \sum_{\nu=1}^\infty\frac{\zeta_\nu}{1+\lambda\cdot \lambda^{-1}_\nu}\phi_\nu(\cdot).
\end{equation*}
Denote $\bar{\zeta}_\nu = \zeta_\nu/(1+\lambda\cdot \lambda^{-1}_\nu)$.
The following lemma gives a non-asymptotic result for the deterministic error.
\begin{lemma}
\label{lem:anormbarf}
For any $n\geq 1$,
\begin{equation*}
\|\bar{\zeta}_{\infty\lambda} - \zeta\|_a \leq \begin{cases}
\frac{1}{2}(1+a)^{(1+a)/2}(1-a)^{(1-a)/2}\lambda^{(1-a)/2}\|\zeta\|_\HH, & \text{if } 0\leq a<1,\\
\|\zeta\|_\HH, & \text{if } a=1.
\end{cases}
\end{equation*}
\end{lemma}
\begin{proof}
For any $0\leq a\leq 1$, we have
\begin{equation*}
\begin{aligned}
\|\bar{\zeta}_{\infty\lambda} - \zeta\|_a^2 & = \sum_{\nu=1}^\infty(1+\lambda_\nu^{-a})(\bar{\zeta}_\nu - \zeta_\nu)^2\\
& \leq \lambda^2\sup_{\nu\geq 1}\frac{(1+\lambda_\nu^{-a})\lambda^{-1}_\nu}{(1+\lambda\cdot\lambda^{-1}_\nu)^2}\sum_{\nu=1}^\infty\lambda^{-1}_\nu \zeta_\nu^2\\
& = \lambda^2\|\zeta\|_{\HH}^2\sup_{\nu\geq 1}\frac{(1+\lambda_\nu^{-a})\lambda^{-1}_\nu}{(1+\lambda\cdot\lambda^{-1}_\nu)^2}.
\end{aligned}
\end{equation*}
Observe that
\begin{equation*}
\sup_{\nu\geq 1}\frac{(1+\lambda_\nu^{-a})\lambda^{-1}_\nu}{(1+\lambda\cdot\lambda^{-1}_\nu)^2} \leq \sup_{x>0}\frac{(1+x^{-a})x^{-1}}{(1+\lambda x^{-1})^2} \leq
\begin{cases}
\frac{(1+a)^{1+a}(1-a)^{1-a}}{2\lambda^{a+1}}, & \text{if } 0\leq a<1,\\
\lambda^{-2}, & \text{if } a=1.
\end{cases}
\end{equation*}
This completes the proof.
\end{proof}

\paragraph{Step 2: Stochastic error}

For any $g,g_1,g_2\in\HH$,  we have the first- and second-order Fr\'echet derivatives  as follows:
\begin{equation*}
\begin{aligned}
&Dl_{n}(g)g_1  = -\frac{1}{n}\sum_{i=1}^n [Y_i - g(X_i)]g_1(X_i),\quad Dl_\infty(g)g_1 = -\langle \zeta-g, g_1\rangle_{L_2(\Pi)}\\
&D^2l_{n}(g)g_1g_2  = \frac{1}{n}\sum_{i=1}^ng_1(X_i)g_2(X_i), \quad D^2l_\infty(g)g_1g_2 =  \langle g_1, g_2\rangle_{L_2(\Pi)}.\\
\end{aligned}
\end{equation*}
Here, the Fr\'echet derivatives are defined in $\|\cdot\|$-norm. Since $D^2l_{\infty\lambda}(\bar{\zeta})\phi_\nu\phi_\mu =\langle D^2l_{\infty\lambda}(\bar{\zeta})\phi_\nu, \phi_\mu\rangle= (1+\lambda\cdot\lambda^{-1}_\nu)\delta_{\nu\mu}$ and $\|\phi_\nu\| = (1+\lambda^{-1}_\nu)^{1/2}$, we have
\begin{equation}
\label{eqn:defdtlinftylambdainv}
\left[D^2l_{\infty\lambda}(\bar{\zeta})\right]^{-1}\phi_\nu = (1+\lambda\cdot\lambda^{-1}_\nu)^{-1}(1+\lambda^{-1}_\nu)\phi_\nu.
\end{equation}
 Define that
\begin{equation*}
\tilde{\zeta}^\dagger \equiv \bar{\zeta}_{\infty\lambda} - \left[D^2l_{\infty\lambda}(\bar{\zeta}_{\infty\lambda})\right]^{-1}Dl_{n\lambda}(\bar{\zeta}_{\infty\lambda}),
\end{equation*}
then we can decompose the stochastic error as follows:
\begin{equation}
\label{eqn:decomstocsterm}
\widehat{\zeta}_{n\lambda} - \bar{\zeta}_{\infty\lambda} = (\tilde{\zeta}^\dagger - \bar{\zeta}_{\infty\lambda}) +  (\widehat{\zeta}_{n\lambda} - \tilde{\zeta}^\dagger).
\end{equation}
We study the two terms on the right-hand side of (\ref{eqn:decomstocsterm}) separately. For  simplicity of the notations, we  abbreviate the subscripts of $\widehat{\zeta}_{n\lambda}$ and $\bar{\zeta}_{\infty\lambda}$ in the rest of this section.

For any $0\leq a\leq 1$ and $\lambda>0$, we define $\Delta(a,\lambda)$ by satisfying
\begin{equation}
\label{eqn:deltaa}
\Delta(a,\lambda)\geq \lambda^{a+d/2m}\sum_{\nu=1}^\infty(1+\lambda_\nu^{-a})(1+\lambda\cdot\lambda^{-1}_\nu)^{-2}.
\end{equation}
Under Assumption \ref{asp:decayrateofeigen}, it is easy to see that there exists a $\Delta(a,\lambda)<\infty$.
The following lemma gives a non-asymptotic bound for  $(\tilde{\zeta}^\dagger - \bar{\zeta})$ in (\ref{eqn:decomstocsterm}).
\begin{lemma}
\label{lem:boundontildefbarf}
For any $0\leq a\leq 1$ and $n\geq 1$, we have that with probability at least $1-3\exp(-\alpha^2)$,
\begin{equation*}
\|\tilde{\zeta}^\dagger - \bar{\zeta}\|_a^2  \leq  \left\{\alpha A\|\zeta\|_\HH + c_\phi\sigma(1+\sqrt{2}\alpha)\right\}^2\Delta(a,\lambda)n^{-1}\lambda^{-(a+d/2m)}.
\end{equation*}
Here, $c_\phi$ is define in Assumption \ref{asp:decayrateofeigen}, $\Delta(a,\lambda)$ is defined in (\ref{eqn:deltaa}), and $A$ is a constant to be given in (\ref{eqn:supg1brg2hk}).
\end{lemma}
\begin{proof}
Observe that
\begin{equation*}
Dl_{n\lambda}(\bar{\zeta}) = Dl_{n\lambda}(\bar{\zeta}) - Dl_{\infty\lambda}(\bar{\zeta}) = Dl_n(\bar{\zeta}) - Dl_{\infty}(\bar{\zeta}).
\end{equation*}
Thus,
\begin{equation}
\label{eqn:snlambdaflambdaeps}
\begin{aligned}
&\sup_{\nu\geq 1}|Dl_{n\lambda}(\bar{\zeta})\phi_\nu| \\
&\leq \sup_{\nu\geq1}\left|\frac{1}{n}\sum_{i=1}^n\left\{(\zeta-\bar{\zeta})(X_i)\phi_\nu(X_i) - \E[(\zeta-\bar{\zeta})(X)\phi_\nu(X)]\right\}\right| \\
 & \quad +\sup_{\nu\geq1} \left|\frac{1}{n}\sum_{i=1}^n\varepsilon_i\phi_\nu(X_i)\right|.
\end{aligned}
\end{equation}
We consider the two terms on the right-hand side of (\ref{eqn:snlambdaflambdaeps}) seperately. For the
first term, since Lemma \ref{lem:anormbarf}  implies $\|\zeta - \bar{\zeta}\|\leq \|\zeta\|_{\HH}$, we can apply
 Lemma \ref{lem:conceninequass} in Section \ref{sec:keylemma} by letting $g = \zeta-\bar{\zeta}$ and $t = \alpha A\|\zeta\|_{\HH}$. Then, with
 probability at least $1-2\exp(-\alpha^2)$,
\begin{equation*}
\begin{aligned}
\sup_{\nu\geq1} \left|\frac{1}{n}\sum_{i=1}^n\left[(\zeta-\bar{\zeta})(X_i)\phi_\nu(X_i) - \E\{(\zeta-\bar{\zeta})(X)\phi_\nu(X)\}\right]\right| \leq \frac{\alpha A\|\zeta\|_\HH}{\sqrt{n}}.
\end{aligned}
\end{equation*}
Then we consider the second term  in (\ref{eqn:snlambdaflambdaeps}). Let $\Sigma_{\phi_\nu} = [\phi_\nu(X_i)\phi_\nu(X_j)]_{1\leq i,j\leq n}$ and $\bsvarep = (\varepsilon_1,\ldots,\varepsilon_n)^\top$. The uniform version of the Hanson-Wright inequality for
suprema of quadratic forms (see, e.g., \cite{talagrand1996new}) gives
\begin{equation*}
\begin{aligned}
 & \P\left(\sup_{\nu\geq1}\sigma^{-2}\bsvarep^\top\Sigma_{\phi_\nu}\bsvarep > \sup_{\nu\geq1}\left\{\text{tr}(\Sigma_{\phi_\nu})+2\sqrt{\text{tr}(\Sigma_{\phi_\nu}^2)}\alpha + 2\|\Sigma_{\phi_\nu}\|\alpha^2\right\}\right)  \\
 & \leq \exp(-\alpha^2).
\end{aligned}
\end{equation*}
Observe that $\text{tr}(\Sigma_{\phi_\nu}) = \sum_{i=1}^n\phi_\nu^2(X_i) \leq nc_\phi^2$ and
\begin{equation*}
\|\Sigma_{\phi_\nu}\|\leq \sqrt{\text{tr}(\Sigma_{\phi_\nu}^2)} = \sqrt{\sum_{i,j=1}^n\phi^2_\nu(X_i)\phi_\nu^2(X_j)}\leq nc_\phi^2,
\end{equation*}
we have probability at least $1-\exp(-\alpha^2)$ such that
\begin{equation*}
\sup_{\nu\geq1}\left|\frac{1}{n}\sum_{i=1}^n\varepsilon_i\phi_\nu(X_i)\right| \leq \frac{c_\phi\sigma(1+\sqrt{2}\alpha)}{\sqrt{n}}.
\end{equation*}
Therefore, (\ref{eqn:snlambdaflambdaeps}) implies that with probability at least $1-3\exp(-\alpha^2)$,
\begin{equation}
\label{eqn:snlambdaflambda}
\begin{aligned}
 \sup_{\nu\geq1}|Dl_{n\lambda}(\bar{\zeta})\phi_\nu| & \leq \alpha A\|\zeta\|_\HH n^{-1/2} + c_\phi\sigma(1+\sqrt{2}\alpha)n^{-1/2}.
\end{aligned}
\end{equation}
By the definition of $\tilde{\zeta}^\dagger$ and (\ref{eqn:defdtlinftylambdainv}),  with probability at least $1-3\exp(-\alpha^2)$,
\begin{equation*}
\begin{aligned}
\|\tilde{\zeta}^\dagger - \bar{\zeta}\|_a^2 & = \|[D^2l_{\infty\lambda}(\bar{\zeta})]^{-1}Dl_{n\lambda}(\bar{\zeta})\|_a^2\\
& = \sum_{\nu=1}^\infty(1+\lambda_\nu^{-a})(1+\lambda\cdot\lambda^{-1}_\nu)^{-2}[Dl_{n\lambda}(\bar{\zeta})\phi_\nu]^2\\
& \leq \left\{\sup_\nu[Dl_{n\lambda}(\bar{\zeta})\phi_\nu]^2\right\}\sum_{\nu=1}^\infty(1+\lambda_\nu^{-a})(1+\lambda\cdot\lambda^{-1}_\nu)^{-2}\\
& \leq \frac{[\alpha A\|\zeta\|_\HH+c_\phi\sigma(1+\sqrt{2}\alpha)]^2}{n}\sum_{\nu=1}^\infty(1+\lambda_\nu^{-a})(1+\lambda\cdot\lambda^{-1}_\nu)^{-2}\\
& \leq \frac{[\alpha A\|\zeta\|_\HH+c_\phi\sigma(1+\sqrt{2}\alpha)]^2}{n}\Delta(a,\lambda) \lambda^{-(a+d/2m)},
\end{aligned}
\end{equation*}
This completes the proof for Lemma \ref{lem:boundontildefbarf}.
\end{proof}

 For any $0<b\leq 1$, we define $c_b$ as
\begin{equation}
\label{eqn:bdongammajb}
c_b \equiv \sum_{\nu=1}^\infty(1+\lambda_\nu^{-b})^{-1}.
\end{equation}
Then, $c_b<\infty$ by Assumption \ref{asp:decayrateofeigen}.
Now we give a non-asymptotic bound for  $(\widehat{\zeta} - \tilde{\zeta}^\dagger)$ in (\ref{eqn:decomstocsterm}).
\begin{lemma}
\label{lem:bdforhatftildef}
For any $\rho>0$ and $\alpha>0$, if there exists some $b\in(d/2m, 1]$ such that
\begin{equation}
\label{eqn:nlambdabd2m}
n^{-1}\lambda^{-(b+d/2m)} < \frac{\rho^2}{\Delta(b,\lambda)\{2\alpha^2c_\phi^4c_b + [\alpha A\|\zeta\|_\HH+c_\phi\sigma(1+\sqrt{2}\alpha)]^2\}},
\end{equation}
 then with probability at least $1-5\exp(-\alpha^2)$,
\begin{equation*}
\|\widehat{\zeta} - \tilde{\zeta}^\dagger\|_a^2 <\frac{2\alpha^2c_\phi^4c_b\Delta(a,\lambda)\rho^2}{(1-\rho)^{2}}n^{-1}\lambda^{-(a+d/2m)}, \ \ \forall 0\leq a\leq 1.
\end{equation*}
Here,  $c_\phi$ is define in Assumption \ref{asp:decayrateofeigen}, $\Delta(a,\lambda)$ is defined in (\ref{eqn:deltaa}), $c_b$ is defined in (\ref{eqn:bdongammajb}), and $A$ is a constant to be given in (\ref{eqn:supg1brg2hk}).
\end{lemma}
\begin{proof}
By the definition of $\tilde{\zeta}^\dagger$, we have that
\begin{equation*}
Dl_{n\lambda}(\bar{\zeta}) = D^2l_{\infty\lambda}(\bar{\zeta})(\bar{\zeta} - \tilde{\zeta}^\dagger).
\end{equation*}
Since $l_{n\lambda}$ is quadratic,
\begin{equation*}
Dl_{n\lambda}(\widehat{\zeta}) = Dl_{n\lambda}(\bar{\zeta}) + D^2l_{n\lambda}(\bar{\zeta})(\widehat{\zeta} - \bar{\zeta}) = 0.
\end{equation*}
Thus,
\begin{equation*}
D^2l_{\infty\lambda}(\bar{\zeta})(\widehat{\zeta} - \tilde{\zeta}^\dagger) = D^2l_\infty(\bar{\zeta})(\widehat{\zeta} - \bar{\zeta}) - D^2l_n(\bar{\zeta})(\widehat{\zeta} - \bar{\zeta}),
\end{equation*}
and this implies
\begin{equation*}
\widehat{\zeta} - \tilde{\zeta}^\dagger = [D^2l_{\infty\lambda}(\bar{\zeta})]^{-1}(D^2l_{\infty}(\bar{f}) - D^2l_n(\bar{\zeta}))(\widehat{\zeta} - \bar{\zeta}).
\end{equation*}
Recall  that$\bar{\zeta} = \sum_{\nu=1}^\infty\bar{\zeta}_\nu\phi_\nu$. We denote $\widehat{\zeta} = \sum_{\nu=1}^\infty\widehat{\zeta}_\nu\phi_\nu$. By the Cauchy-Schwarz inequality,
\begin{equation*}
\begin{aligned}
\|\widehat{\zeta} - \tilde{\zeta}^\dagger\|_a^2 & = \sum_{\nu=1}^\infty(1+\lambda_\nu^{-a})(1+\lambda\cdot\lambda_\nu^{-1})^{-2}\\
& \quad\quad\times\left[\sum_{j=1}^\infty(\widehat{\zeta}_j - \bar{\zeta}_j)\left(\frac{1}{n}\sum_{i=1}^n\phi_j(X_i)\phi_\nu(X_i) - \langle \phi_j,\phi_\nu\rangle_{L_2(\Pi)}\right)\right]^2\\
& \leq \sum_{\nu=1}^\infty(1+\lambda_\nu^{-a})(1+\lambda\cdot\lambda^{-1}_\nu)^{-2}\left[\sum_{j=1}^\infty(\widehat{\zeta}_j - \bar{\zeta}_j)^2(1+\lambda_j^{-b})\right]\\
& \quad\quad\times\left(\sum_{j=1}^\infty(1+\lambda_j^{-b})^{-1}\left[\frac{1}{n}\sum_{i=1}^n\phi_j(X_i)\phi_\nu(X_i) - \langle \phi_j,\phi_\nu\rangle_{L_2(\Pi)}\right]^2\right).
\end{aligned}
\end{equation*}
Since $|\phi_j(X_i)\phi_\nu(X_i)|\leq c_\phi^2$ and $\phi_j(\cdot)\phi_\nu(\cdot)$ is measurable, by applying the McDiarmid's  inequality, we have that with probability at least $1-2\exp(-\alpha^2)$,
\begin{equation}
\label{eqn:bdonphijphinu}
\sup_{j,\nu\geq1}\left[\frac{1}{n}\sum_{i=1}^n\phi_j(X_i)\phi_\nu(X_i) - \langle\phi_j,\phi_\nu\rangle_{L_2(\Pi)}\right]^2 \leq \frac{2\alpha^2 c_\phi^4}{n}.
\end{equation}

Combine  (\ref{eqn:deltaa}), (\ref{eqn:bdongammajb}), and (\ref{eqn:bdonphijphinu}), then for any $d/2m<b\leq 1$, we have probability at least $1-2\exp(-\alpha^2)$,
\begin{equation}
\label{eqn:anormhatftildef}
\|\widehat{\zeta} - \tilde{\zeta}^\dagger\|_a^2 \leq \frac{2\alpha^2c_\phi^4\Delta(a,\lambda)c_b}{n}\lambda^{-(a+d/2m)}\|\widehat{\zeta} - \bar{\zeta}\|_b^2.
\end{equation}
Take $a=b$, then with probability at least $1-2\exp(-\alpha^2)$,
\begin{equation*}
\|\widehat{\zeta} - \tilde{\zeta}^\dagger\|_b^2 \leq \frac{2\alpha^2c_\phi^4\Delta(b,\lambda)c_b}{n}\lambda^{-(b+d/2m)}\|\widehat{\zeta} - \bar{\zeta}\|_b^2.
\end{equation*}
If (\ref{eqn:nlambdabd2m}) holds,
then $\|\widehat{\zeta} - \tilde{\zeta}\|_b< \rho \|\widehat{\zeta} - \bar{\zeta}\|_b$ and $\|\tilde{\zeta} - \bar{\zeta}\|_b \geq \|\widehat{\zeta} - \bar{\zeta}\|_b - \|\widehat{\zeta} - \tilde{\zeta}\|_b> (1-\rho) \|\widehat{\zeta} - \bar{\zeta}\|_b$. By Lemma \ref{lem:boundontildefbarf}, with probability at least $1-5\exp(-\alpha^2)$,
\begin{equation*}
\begin{aligned}
\|\widehat{\zeta} - \bar{\zeta}\|^2_b & < \frac{[\alpha A\|\zeta\|_\HH+c_\phi\sigma(1+\sqrt{2}\alpha)]^2\Delta(b)}{(1-\delta)^2n}\lambda^{-(b+d/2m)} < \frac{\rho^2}{(1-\rho)^2}.
\end{aligned}
\end{equation*}
where the second inequality is from (\ref{eqn:nlambdabd2m}). We complete the proof by plugging the above inequality to (\ref{eqn:anormhatftildef}).
\end{proof}

\paragraph{Step 3: Putting it together}
We consider for $a=0$. Let
\begin{equation*}
\Delta(0,\lambda) = \frac{4m}{4m-d}C_\lambda^{d/2m},
\end{equation*}
which satisfies the definition (\ref{eqn:deltaa}) and it does not depend on $\lambda$.
Let
\begin{equation}
\label{eqn:defoflambdadelatao}
\begin{aligned}
\lambda & = n^{-\frac{2m}{2m+d}}\left\{2\sqrt{\Delta(0,\lambda)}\left[\alpha A + \frac{\sigma c_\phi (1+\sqrt{2}\alpha)}{\|\zeta\|_\HH}\right]\right\}^{\frac{4m}{2m+d}}\\
& = n^{-\frac{2m}{2m+d}}\left\{2\sqrt{\frac{4m}{4m-d}}C_\lambda^{d/4m}\left[\alpha A + \frac{\sigma c_\phi (1+\sqrt{2}\alpha)}{\|\zeta\|_\HH}\right]\right\}^{\frac{4m}{2m+d}}
\end{aligned}
\end{equation}
Then for any $b\in(d/2m,1)$ and $\rho$ satisfying
\begin{equation*}
\rho>c_\rho n^{-\frac{m(1-b)}{2m+d}}\alpha^{\frac{2m(1-b)}{2m+d}}\left(1+\frac{\sigma}{\|\zeta\|_\HH}\right)^{-\frac{2mb+d}{2m+d}}(\|\zeta\|_\HH+\sigma),
\end{equation*}
 the $\lambda$ defined in (\ref{eqn:defoflambdadelatao}) satisfies the condition (\ref{eqn:nlambdabd2m}) in Lemma \ref{lem:bdforhatftildef}. Thus, Lemma \ref{lem:bdforhatftildef} implies
\begin{equation}
\|\widehat{\zeta} - \tilde{\zeta}^\dagger\|_{L_2(\Pi)}\leq \tilde{c} \alpha^{\frac{4m-d}{2m+d}}n^{-\frac{4m-d}{4m+2d}}(\|\zeta\|_\HH+\sigma)\left(1+\frac{\sigma}{\|\zeta\|_\HH}\right)^{-\frac{3d}{2m+d}},
\end{equation}
for some constant $\tilde{c}$ not depending on $n,  \sigma, m, d, \|\zeta\|_\HH$.
Combining Lemma \ref{lem:anormbarf},  \ref{lem:boundontildefbarf} and  \ref{lem:bdforhatftildef},  with probability at least $1-8\exp(-\alpha^2)$,
\begin{equation*}
\begin{aligned}
& \|\widehat{\zeta} - \zeta\|_{L_2(\Pi)} \leq C_2\left[1+\alpha^{\frac{2m-d}{2m+d}}n^{-\frac{2m-d}{4m+2d}}\left(1+\frac{\sigma}{\|\zeta\|_\HH}\right)^{-\frac{2d}{2m+d}}\right]\\
& \quad\quad\quad\quad\quad\quad\quad\quad\quad\cdot\alpha^{\frac{2m}{2m+d}}n^{-\frac{m}{2m+d}}(\|\zeta\|_\HH+\sigma)\left(1+\frac{\sigma}{\|\zeta\|_\HH}\right)^{-\frac{d}{2m+d}},
\end{aligned}
\end{equation*}
for some constant $C_2$ not depending on $n, \sigma, m, d, \|\zeta\|_\HH$ and $\lambda$ given by (\ref{eqn:defoflambdadelatao}). This completes the proof of Theorem \ref{thm:hatfferror}.

\subsubsection{The optimal calibration result: Theorem \ref{thm:thetastardeltahh}}

By Assumption \ref{asp:etaspace}, it is known $\delta(\cdot,\theta) = \zeta(\cdot) - \eta(\cdot,\theta)\in\HH$.
Similar to Theorem \ref{thm:lowerbdnonasymnonpara}, we have that for any $\theta\in\Theta$ and $n\geq 1$,
 \begin{equation}
 \label{eqn:minlowbdtildedelta}
\begin{aligned}
&  \inf_{\widetilde{\delta}}\sup_{\delta(\cdot,\theta)\in\HH}\P\left\{\|\widetilde{\delta}(\cdot,\theta)- \delta(\cdot,\theta)\|_{L_2(\Pi)}^2 \geq C_1' n^{-\frac{2m}{2m+d}}\cdot\right.\\
&\quad\quad\quad\quad\quad\quad\quad\left.\left(\|\delta(\cdot,\theta)\|_\HH+\sigma\right)^2\left(1+{\sigma}/{\|\delta(\cdot,\theta)\|_\HH}\right)^{-\frac{2d}{2m+d}}\right\}>0,
\end{aligned}
\end{equation}
where the constant $C_1>0$ is the same as in Theorem \ref{thm:lowerbdnonasymnonpara} that does not depend on $n, \sigma, m, d, \|\delta(\cdot,\theta)\|_\HH$.
The proof of (\ref{eqn:minlowbdtildedelta}) is identical to the proof of Theorem \ref{thm:lowerbdnonasymnonpara} by replacing $\zeta(\cdot)$ with $\delta(\cdot,\theta)$ and we omit the details. This also proves the Corollary \ref{cor:lowerbd}.

On the other hand, similar to Theorem \ref{thm:hatfferror}, we have that for any $\theta\in\Theta$, $n\geq 1$, and $\alpha>0$, with probability at least $1-8e^{-\alpha^2}$,
\begin{equation}
\label{eqn:hatdeltadeltal2pi}
\begin{aligned}
&  \min_{\lambda>0}\|\delta(\cdot,\theta)-\widehat{\delta}_{n\lambda}(\cdot,\theta)\|_{L_2(\Pi)}^2 \\
&\leq C_2\left[1+\alpha^{\frac{2m-d}{2m+d}}n^{-\frac{2m-d}{4m+2d}}\left(1+{\sigma}/{\|\delta(\cdot,\theta)\|_\HH}\right)^{-\frac{2d}{2m+d}}\right]^2\\
& \quad\quad\cdot\alpha^{\frac{4m}{2m+d}}n^{-\frac{2m}{2m+d}}\left(\|\delta(\cdot,\theta)\|_\HH+\sigma\right)^2\left(1+{\sigma}/{\|\delta(\cdot,\theta)\|_\HH}\right)^{-\frac{2d}{2m+d}},
\end{aligned}
\end{equation}
where the $C_2>0$ is the same as in Theorem  \ref{thm:hatfferror} that does not depend on $n, \sigma, \|\delta(\cdot,\theta)\|_\HH$, and the estimator $\widehat{\delta}_{n\lambda}(\cdot,\theta)$ is defined in (\ref{eqn:widehatdeltanlambda}).
The proof of (\ref{eqn:hatdeltadeltal2pi}) is identical to the proof of Theorem \ref{thm:hatfferror} only by replacing $\zeta(\cdot)$ with $\delta(\cdot,\theta)$  and we also omit the details. This also proves the Corollary \ref{cor:predoptcal}.

Combining (\ref{eqn:minlowbdtildedelta}) and (\ref{eqn:hatdeltadeltal2pi}), then for any $\theta\in\Theta$ and $n\geq 1$, the minimax optimal risk for estimating $\delta(\cdot,\theta)=\zeta(\cdot) - \eta(\cdot,\theta)\in\HH$ is
\begin{equation}
\label{eqn:minimaxoptdelta}
\begin{aligned}
&\|\widetilde{\delta}(\cdot,\theta)- \delta(\cdot,\theta)\|_{L_2(\Pi)}^2 \\
&= C' n^{-\frac{2m}{2m+d}}\left(\|\delta(\cdot,\theta)\|_\HH+\sigma\right)^2\left(1+{\sigma}/{\|\delta(\cdot,\theta)\|_\HH}\right)^{-\frac{2d}{2m+d}},
\end{aligned}
\end{equation}
for some constant $C'$.
Thus, by the definition of $\theta^{\text{opt-pred}}$ in (\ref{def:btheta0}) and (\ref{eqn:minimaxoptdelta}), \begin{equation*}
\begin{aligned}
\theta^{\text{opt-pred}}  & =  \underset{\theta\in\Theta}{\arg\min} \left\{\inf_{\widetilde{\delta}}\|\delta(\cdot,\theta) - \widetilde{\delta}(\cdot,\theta)\|_{L_2(\Pi)}\right\}\\
& = \underset{\theta\in\Theta}{\arg\min} \left\{ C' n^{-\frac{2m}{2m+d}}\left(\|\delta(\cdot,\theta)\|_\HH+\sigma\right)^2\left(1+{\sigma}/{\|\delta(\cdot,\theta)\|_\HH}\right)^{-\frac{2d}{2m+d}}\right\}\\
& = \underset{\theta\in\Theta}{\arg\min} \left\{\|\delta(\cdot,\theta)\|_\HH\right\}\\
& = \underset{\theta\in\Theta}{\arg\min} \left\{\|\zeta(\cdot) - \eta(\cdot,\theta)\|_\HH\right\},
\end{aligned}
\end{equation*}
where the third step is by the fact that parameters $n, \sigma, m, d$ are fixed and
the right-hand side of (\ref{eqn:minimaxoptdelta})  is monotonically increasing as the RKHS norm $\|\delta(\cdot,\theta)\|_\HH$ increases,  and the last step above is by the definition of $\delta(\cdot,\theta)$.
This completes the proof.

\subsubsection{Improvement of prediction by computer models: Theorem \ref{thm:zetahhles}}

By comparing Theorem \ref{thm:hatfferror} with Corollary \ref{cor:predoptcal}, we know that for any finite sample size $n\geq 1$, $\min_{\lambda>0}\|\zeta_{n\lambda}^{\text{opt-pred}}(\cdot)- \zeta(\cdot)\|_{L_2(\Pi)}^2$ has a smaller upper bound than $\min_{\lambda>0}\|\widehat{\zeta}_{n\lambda}(\cdot)- \zeta(\cdot)\|_{L_2(\Pi)}^2$. 
This completes the proof.

\subsubsection{Comparison with frequentist calibrations: Remark \ref{prop:l2calibrationpred}}
For the calibration part,
\cite{tuo2015efficient}  shows 
$\widehat{\theta}_n^{L_2} \to_\P \theta^{L_2}$, which implies that 
\begin{equation*}
\P\{\widehat{\theta}_n^{L_2} \to \theta^{L_2}\}>0.
\end{equation*}
On the other hand,  it is known that  $\widehat{\theta}_n^{l_2}$ converges to the minimizer of Kullback-Leibler distance between $\zeta(\cdot)$ and $\eta(\cdot,\theta)$ (see, e.g., \cite{white1982maximum}), that is,
$\widehat{\theta}_n^{l_2}\to_{a.s.}\theta^{L_2}$, which also implies that
\begin{equation*}
\P\{\widehat{\theta}_n^{l_2} \to \theta^{L_2}\}>0.
\end{equation*}

For the prediction part, observe that
\begin{equation*}
\begin{aligned}
\|\eta(\cdot,\widehat{\theta}_n^{L_2})- \zeta(\cdot)\|_{L_2(\Pi)}^2 & \geq \|\eta(\cdot,\theta^{L_2})- \zeta(\cdot)\|_{L_2(\Pi)}^2\\
& \geq \min_{\lambda>0}\|\eta(\cdot,\theta^{L_2})+\widehat{\delta}_{n\lambda}(\cdot,\theta^{L_2})- \zeta(\cdot)\|_{L_2(\Pi)}^2,
\end{aligned}
\end{equation*}
where the first step is by definition of $\theta^{L_2}$ and the second step is due to the fact that when $\lambda\to\infty$, $\widehat{\delta}_{n\lambda}\to 0$ (see, e.g., \cite{wahba1990}).
On the other hand
\begin{equation*}
\min\left\{\|\delta(\cdot, \theta^{L_2})\|_{\HH}, \|\delta(\cdot, \widehat{\theta}_n^{l_2})\|_{\HH}\right\}\geq\|\delta(\cdot, \theta^{\text{opt-pred}})\|_{\HH} = \min_{\theta\in\Theta}\|\delta(\cdot,\theta)\|_\HH.
\end{equation*}
By Corollary \ref{cor:predoptcal}, with probability approaching to one, 
$\min_{\lambda>0}\|\eta(\cdot,\theta^{\text{opt-pred}})+\widehat{\delta}_{n\lambda}(\cdot,\theta^{\text{opt-pred}})- \zeta(\cdot)\|_{L_2(\Pi)}^2$ achieves a smaller upper bound  of the predictive mean squared error than 
$\min_{\lambda>0}\|\eta(\cdot,\theta^{L_2})+\widehat{\delta}_{n\lambda}(\cdot,\theta^{L_2})- \zeta(\cdot)\|_{L_2(\Pi)}^2$ and $\min_{\lambda>0}\|[\eta(\cdot,\widehat{\theta}_n^{l_2})+\widehat{\delta}_{n\lambda}(\cdot,\widehat{\theta}_n^{l_2})]- \zeta(\cdot)\|_{L_2(\Pi)}^2$. This completes the proof.

\subsection{Proof for Section \ref{sec:algorithm}}
\label{sec:proofsec4}

\subsubsection{Connection with Bayesian calibration: Proposition \ref{prop:equivalencewithbayesian}}
Let $T$ be the $n\times n$ matrix with $ij$th entry $h_j(X_i)$.
Under the prior distribution  (\ref{eqn:bayesianmodel}), 
\begin{equation*}
\begin{aligned}
\E[\vec{Y}\vec{Y}^\top] & = \alpha TT^\top+\beta \Sigma+\sigma^2I,\\
\E[\zeta(x)\vec{Y}] & = \alpha T\left( \begin{array}{c}
h_1(x,\theta) \\
\vdots  \\
h_M(x,\theta)  \end{array} \right) + \beta\left( \begin{array}{c}
K(X_1,x) \\
\vdots  \\
K(X_n,x)  \end{array} \right).
\end{aligned}
\end{equation*}
Then, with $\lambda=\sigma^2/n\beta$, 
\begin{equation*}
\begin{aligned}
\E[\zeta(x)|\vec{Y}] & = \{\E[\zeta(x)\vec{Y}]\}^\top\{\E[\vec{Y}\vec{Y}^\top]\}^{-1}\vec{Y} \\
& =  \left(h_1(x),\ldots,h_M(x)\right)\alpha\beta^{-1} T^\top\left(\alpha \beta^{-1}TT^\top+\Sigma+n\lambda I\right)^{-1}\vec{Y}  \\
&  \quad + \left(K(X_1,x),\ldots,K(X_n,x)\right)\left(\alpha\beta^{-1}TT^\top+\Sigma+n\lambda I\right)^{-1}\vec{Y}.
\end{aligned}
\end{equation*}
Note that 
\begin{equation*}
\begin{aligned}
&\lim_{\alpha\to\infty}(\alpha\beta^{-1}TT^\top+\Sigma+n\lambda I)^{-1} \\
& \ \  = (\Sigma+n\lambda I)^{-1}\{I-T(T^\top (\Sigma+n\lambda I)^{-1}T)^{-1}T^\top (\Sigma+n\lambda I)^{-1}\},
\end{aligned}
\end{equation*}
and
\begin{equation*}
\begin{aligned}
&\lim_{\alpha\to\infty}\alpha\beta^{-1}T^\top(\alpha\beta^{-1}TT^\top+\Sigma+n\lambda I)^{-1} \\
& \ \  = \{T^\top (\Sigma+n\lambda I)^{-1}T\}^{-1}T^\top (\Sigma+n\lambda I)^{-1},
\end{aligned}
\end{equation*}
which are the solution to the smoothing splines (see, e.g., \cite{wahba1990}).
Therefore,
\begin{equation*}
\begin{aligned}
\lim_{a\to\infty}\widehat{\zeta}_a(x)  & = \left(h_1(x),\ldots,h_M(x)\right) \{T^\top (\Sigma+n\lambda I)^{-1}T\}^{-1}T^\top (\Sigma+n\lambda I)^{-1}\vec{Y}\\
&  \quad+ \left(K(X_1,x),\ldots,K(X_n,x)\right)(\Sigma+n\lambda I)^{-1}\\
& \quad\quad\quad\cdot \{I-T(T^\top (\Sigma+n\lambda I)^{-1}T)^{-1}T^\top (\Sigma+n\lambda I)^{-1}\}\vec{Y}\\
& = \left(h_1(x),\ldots,h_M(x)\right)\widehat{\theta}^{\text{opt-pred}} + \widehat{\delta}_{n\lambda}(\cdot)= \widehat{\zeta}_{n\lambda}^{\text{opt-pred}}(x).
\end{aligned}
\end{equation*}
This completes the proof.

\subsection{Key lemma}
\label{sec:keylemma}

\begin{lemma}
\label{lem:conceninequass}
Recall that $R(\cdot,\cdot)$ is the reproducing kernel associated with  $(\HH,\|\cdot\|)$.
Suppose  that $c_{R} = \sup_{x\in\Omega}\sqrt{R(x,x)}$ is finite.
Then for any $t\geq 0$ and $\nu\geq 1$, we have
\begin{equation*}
\begin{aligned}
& \P\left(\sup_{\underset{g:\|g\|\leq \|\zeta\|_\HH}{\nu\geq1}}\left|\frac{1}{\sqrt{n}}\sum_{i=1}^n\left\{g(X_i)\phi_\nu(X_i) - \E[g(X)\phi_\nu(X)]\right\}\right|\geq t\right)\\
 &\leq 2\exp\left(-\frac{t^2}{A^2\|\zeta\|_\HH^{2}}\right).
\end{aligned}
\end{equation*}
Here, $A$ is a constant  given in (\ref{eqn:supg1brg2hk}) and it does not depend on $n,\sigma,\|\zeta\|_\HH$.
\end{lemma}

\begin{proof}
For any $g_1, g_2$ satisfying $\|g_1\|\leq\|\zeta\|_\HH, \|g_2\|\leq\|\zeta\|_\HH$, we have that for any $\nu\geq 1$,
\begin{equation*}
|g_1(X_i)\phi_\nu(X_i) - g_2(X_i)\phi_\nu(X_i)| \leq \max_{x\in\Omega}|g_1(x)-g_2(x)|c_{\phi},
\end{equation*}
where $c_\phi$ is defined in Assumption \ref{asp:decayrateofeigen}.
Let
\begin{equation*}
Z_n(g,\phi_\nu) = \frac{1}{\sqrt{n}}\sum_{i=1}^n\left[g(X_i)\phi_\nu(X_i) - \E\{g(X)\phi_\nu(X)\}\right].
\end{equation*}
Since $\Pi(\Omega) = 1$, by the Azuma-Hoeffding inequality, we have for any $t>0$,
\begin{equation}
\label{eqn:pfnznfg1}
\begin{aligned}
&\P(|Z_{n}(g_1,\phi_\nu) - Z_{n}(g_2,\phi_\nu)|\geq t)\\
&\quad\leq 2\exp\left(-\frac{t^2}{8c_\phi^2\max_{x\in\Omega}|g_1(x)-g_2(x)|^2}\right).
\end{aligned}
\end{equation}
In the following, we apply the maximal inequalities for empirical process. See, e.g., \cite{kosorok2007introduction}.
Recall that the Orlics norm $\|Z\|_{\psi_2}$ for any random variable $Z$ is
\begin{equation*}
\|Z\|_{\psi_2} \equiv \inf_{c>0}\left\{\E\psi_2(|Z|/c)\leq 1\right\},
\end{equation*}
where $\psi_2(x) \equiv \exp(x^2) -1$.  By (\ref{eqn:pfnznfg1}) and Lemma 8.1 in \cite{kosorok2007introduction}, we obtain that
\begin{equation}
\label{eqn:psi2norminfty}
\left\||Z_{n}(g_1,\phi_\nu) - Z_{n}(g_2,\phi_\nu)|\right\|_{\psi_2}\leq \sqrt{24}c_\phi\|g_1-g_2\|_{L_\infty(\Omega)}.
\end{equation}
Let $\tau = \sqrt{\log \frac{3}{2}}$ and $\psi(x) = \psi_2(\tau x)$.
Then, $\psi(\cdot)$ is convex, nondecreasing with $\psi(0)=0$ and $\psi(1)\leq \frac{1}{2}$. Moreover, since $\forall x,y\geq 1$, $\tau^{x^2}(\tau^{x^2(y^2 - 1)}+1-\tau^{y^2})\geq \tau(\tau^{y^2-1}+1-\tau^{y^2})>2-\tau^{y^2}$, we have $\psi(x)\psi(y)\leq \psi(xy)$. From the proof of Lemma 8.2 in \cite{kosorok2007introduction},  for any random variables $Z_1,\ldots,Z_k$,
\begin{equation}
\label{eqn:maxi1mtildexi}
\left\|\max_{1\leq i\leq k}Z_i\right\|_{\psi_2}\leq \frac{2}{\tau}\psi_2^{-1}(k)\max_{1\leq i\leq k}\|Z_i\|_{\psi_2}.
\end{equation}
We define a ball $\BB_{\|\zeta\|_\HH}(\|\cdot\|) = \{g\in\HH:\|g\|\leq \|\zeta\|_\HH\}$.
It is known the covering number of $\BB_{\|\zeta\|_\HH}(\|\cdot\|)$, which we denote by
\begin{equation*}
\NN\left\{\kappa, \BB_{\|\zeta\|_\HH}(\|\cdot\|), \|\cdot\|_{L_\infty(\Omega)}\right\},\quad\text{for any } \kappa>0,
\end{equation*}
has the following bound (see, e.g., \cite{edmunds1996function})
\begin{equation}
\label{eqn:covnumbrhk}
\log \NN\left\{\kappa, \BB_{\|\zeta\|_\HH}(\|\cdot\|), \|\cdot\|_{L_\infty}(\Omega)\right\}\leq c_0\left(\frac{\|\zeta\|_\HH}{\kappa}\right)^{d/{m}}.
\end{equation}
Here, $c_0$ is independent of $\|\zeta\|_\HH$ and $\kappa$.
Note for any $g\in\BB_{\|\zeta\|_\HH}(\|\cdot\|)$ and $x\in\Omega$,  $|g(x)| = |\langle g(\cdot), R(x,\cdot)\rangle|\leq \|g\|\sqrt{R(x,x)}$. Hence, $\|g\|_{L_\infty(\Omega)}=\max_{x\in\Omega}|g(x)|\leq \|\zeta\|_{\HH}\cdot c_{R}$.
By the general maximal inequality (see, e.g., Theorem 8.4 in \cite{kosorok2007introduction}), (\ref{eqn:psi2norminfty}), (\ref{eqn:maxi1mtildexi}) and (\ref{eqn:covnumbrhk}), we obtain that
\begin{equation*}
\left\|\sup_{\underset{g_1,g_2\in\BB_{\|\zeta\|_\HH}(\|\cdot\|), \|g_1-g_2\|_{L_\infty(\Omega)}\leq \|\zeta\|_\HH \cdot c_{R}}{\nu\geq1}}|Z_n(g_1,\phi_\nu) - Z_n(g_2,\phi_\nu)|\right\|_{\psi_2} \leq A\|\zeta\|_\HH,
\end{equation*}
where
\begin{equation}
\label{eqn:supg1brg2hk}
\begin{aligned}
&A=A(c_\phi,c_R,d,m)  \\
& = \frac{32\sqrt{6}c_\phi c_0^{m/d}}{\sqrt{\log 1.5}} \int_0^{c_0^{-m/d}c_R}\sqrt{\log[1+\exp(x^{-d/m})]}dx \\
& \quad+ \frac{40\sqrt{6}c_\phi c_R}{\sqrt{\log 1.5}}\sqrt{\log[1+\exp(2^{1-d/m}c_0c_R^{-d/m})]}.
\end{aligned}
\end{equation}
By letting  $g_2 = 0$ in (\ref{eqn:supg1brg2hk}) and the Lemma 8.1 in \cite{kosorok2007introduction}, we complete the proof for Lemma \ref{lem:conceninequass}.

\end{proof}